\documentclass[a4paper]{article}
\pdfoutput=1
\usepackage{jheppub}

\usepackage[utf8]{inputenc}

\usepackage{amssymb}
\usepackage{stmaryrd}
\usepackage{amsmath}
\usepackage{amsfonts}
\usepackage{mathrsfs}
\usepackage{amsmath,amssymb,amsfonts,amsthm}
\usepackage{slashed}
\usepackage{graphicx}
\usepackage{mathtools}

\usepackage{titlesec}
\titleformat{\paragraph}
{\normalfont\normalsize\centering\itshape}{\theparagraph}{1em}{}
\titlespacing*{\paragraph}
{0pt}{4.5ex plus 1ex minus .2ex}{4.5ex plus .2ex}

\usepackage{hyperref} 

\newtheorem{pro}{Proposition}

\newcommand{\Pl}{\ell_{\mathrm{p}}} 
\newcommand{\sgn}{\mathrm{sgn}} 
\newcommand{\tr}{\mathrm{tr}} 
\newcommand{\dd}{{\rm d}}
\newcommand{\muh}{{\mu_{\rm H}}} 
\newcommand{\g}{\mathbf{g} } 
\newcommand{\s}{{\mathfrak{s} }}
\renewcommand{\t}{{\mathfrak{t}} }

\DeclarePairedDelimiter\floor{\lfloor}{\rfloor}

\title{First-Order Quantum Correction in Coherent State Expectation Value of Loop-Quantum-Gravity Hamiltonian: Overview and Results}

\author[1]{Cong Zhang}  

\affiliation[1]{Faculty of Physics, University of Warsaw, Pasteura 5, 02-093 Warsaw, Poland}

\author[2]{\ Shicong Song}  

\author[2,3]{\ Muxin Han}  

\affiliation[2]{Department of Physics, Florida Atlantic University, 777 Glades Road, Boca Raton, FL 33431-0991, USA}

\affiliation[3]{Institut f\"ur Quantengravitation, Universit\"at Erlangen-N\"urnberg, Staudtstr. 7/B2, 91058 Erlangen, Germany}

\emailAdd{czhang(AT)fuw.edu.pl}
\emailAdd{ssong2019(AT)fau.edu}
\emailAdd{hanm(At)fau.edu}

\abstract{Given the Loop-Quantum-Gravity (LQG) non-graph-changing Hamiltonian $\widehat{H[N]}$, the coherent state expectation value $\langle\widehat{H[N]}\rangle$ admits an semiclassical expansion in $\ell^2_{\rm p}$. In this paper, we compute explicitly the expansion of $\langle\widehat{H[N]}\rangle$ on the cubic graph to the linear order in $\ell^2_{\rm p}$, when the coherent state is peaked at the homogeneous and isotropic data of cosmology. In our computation, a powerful algorithm is developed to overcome the complexity in computing $\langle \widehat{H[N]} \rangle$. In particular, some key innovations in our algorithm substantially reduce the computational complexity in the Lorentzian part of $\langle\widehat{H[N]}\rangle$. Moreover, the algorithm developed in the present work  make it possible to compute the expectation value of arbitrary monomial of holonomies and fluxes on one edge up to arbitrary order of $\Pl^2$.}


\begin{document}

\maketitle

\section{Introduction}\label{sec:intro}

 Loop Quantum Gravity (LQG) is an approach toward the background independent and nonperturbative quantum gravity in four and higher dimensions \cite{thiemann2007modern,rovelli2014covariant,han2007fundamental,ashtekar2004back}. The research on the quantum dynamics of LQG is active and leads to important recent progresses. In particular, there have been tremendous progresses in both canonical and covariant LQG on the semiclassical limit and the consistency with classical gravity e.g. \cite{Giesel:2006uk,giesel2007algebraic,Han:2020chr,Conrady:2008mk,semiclassical,HZ,Bianchi:2006uf,Han:2018fmu}. However, there has been less progress on quantum corrections in the full theory of LQG dynamics (see e.g. \cite{Han:2020fil,Dona:2019dkf} for some results in the covariant approach). It is important that as a candidate of quantum gravity theory, LQG should provide predictions on quantum corrections to the classical theory of gravity.  

The present paper focuses on the canonical LQG.  {In canonical LQG, the dynamics is encoded in the Hamiltonian constraint operator $\widehat{H[N]}=\widehat{H_E[N]}+(1+\beta^2)\widehat{H_L[N]}$, where $N$ is the lapse function, $\widehat{H_E[N]}$ and $\widehat{H_L[N]}$ are the Euclidean part and Lorentzian part of $\widehat{H[N]}$, and $\beta$ is referred to as the Barbero-Immirzi parameter \cite{barbero1995real}.
}
There has been long-term confusion that the quantum dynamics of LQG might not be computable analytically due to the non-polynomial Hamiltonian constraint operator $\widehat{H[N]}$ \cite{Nicolai:2005mc},  {} This confusion has been partially resolved by \cite{giesel2007algebraic}, where the authors schematically show that the coherent state expectation value of the Hamiltonian/master constraint are computable order-by-order by the semiclassical expansion in $\hbar$. The scheme proposed in \cite{giesel2007algebraic} is applicable to a wide class of non-polynomial operators used in studying LQG dynamics. Although this scheme was proposed as early as in 2006 when \cite{giesel2007algebraic} firstly appeared, the expectation value $\langle\widehat{H[N]}\rangle$ has only been computed at the 0-th order (in $\hbar$) which is the classical limit, while the $O(\hbar)$ quantum correction was not studied in the literature, due to the complexity of the operator, especially the complexity of the Lorentzian part $\widehat{H_L[N]}$ in $\widehat{H[N]}$.

The main purpose of our work is to fill this gap by computing explicitly the $O(\hbar)$ quantum correction in the coherent state expectation value $\langle\widehat{H[N]}\rangle$. In this paper, a powerful algorithm is developed to overcome the complexity of $\widehat{H[N]}$, so that we can compute the quantum correction in $\langle\widehat{H[N]}\rangle$, where $\widehat{H[N]}$ is the non-graph-changing Hamiltonian on a cubic lattice $\gamma$. Applying the algorithm, we explicitly expand $\langle\widehat{H[N]}\rangle$ to linear order in $\hbar$, when the coherent state is peaked at the homogeneous and isotropic data of cosmology. Namely we explicitly compute $H_0$ and $H_1$ in
\begin{equation}\label{hamexp0}
\langle\widehat{H[1]}\rangle=H_0+\ell_{\rm p}^2 H_1+O(\ell_{\rm p}^4),\quad \ell_{\rm p}^2=\hbar \kappa
\end{equation} 
where $\kappa=8\pi G_{\rm Newton}$ and the lapse function $N=1$. The 0-th order $H_0$ reproduces the $\mu_0$-scheme cosmological effective Hamiltonian \cite{yang2009alternative,Dapor:2017rwv,han2020effective,Kaminski:2020wbg}. Our new result, $H_1$, gives the first order quantum correction. The explicit expression of $H_1$ is given in Section \ref{sec:results}. Note that in our work, the coherent state for $\langle\widehat{H[N]}\rangle$ is not SU(2) gauge invariant (see below for the motivation). 

This work closely relates to the reduced phase space formulation of LQG (see e.g. \cite{giesel2010algebraicIV,giesel2015scalar}. In this formulation, gravity is coupled to some matter fields known as clock fields. These matter fields serve as material reference frames used to transform gravity variables to gauge invariant Dirac observables. This procedure resolves the Diffeomorphism and Hamiltonian constraints at the classical level and results in the reduced phase space $\mathscr{P}_{red}$ of Dirac observables. The dynamics of the gravity-clock system is described by the material-time evolution generated by the physical Hamiltonian ${\bf H}$ on $\mathscr{P}_{red}$. As an interesting model, we choose the clock fields to be the Gaussian dust \cite{Kuchar:1990vy,giesel2015scalar}. Then the resulting $\mathscr{P}_{red}$ is identified to the pure-gravity unconstrained phase space. This identification defines the pure-gravity Hamiltonian constraint with unit lapse $H[1]$ on $\mathscr{P}_{red}$, and we find ${\bf H}=H[1]$ for gravity coupled to Gaussian dust. In this model, the LQG quantization of $\mathscr{P}_{red}$ is the same as quantizing the pure-gravity unconstrained phase space, and leads to the physical Hilbert space $\mathcal{H}$ identical to the kinematical Hilbert space in the usual LQG. $\mathcal{H}$ is free of constraint because it is from the quantization of $\mathscr{P}_{red}$. The physical Hamiltonian operator is obtained by $\widehat{\bf H}=\frac{1}{2}(\widehat{H[1]}+\widehat{H[1]}^\dagger)$ with the LQG quantization $\widehat{H[N]}$ \cite{thiemann1998quantum,Giesel:2006uj}. Therefore from the reduced phase space LQG perspective, our work computes the expectation value $\langle \widehat{\bf H} \rangle$ at the coherent state peaked at cosmological data on the graph $\gamma$, and gives $\langle \widehat{\bf H} \rangle$ equals the real part of \eqref{hamexp0}.

Recent research works have been focus on building models of LQG on a single graph $\gamma$ \cite{Dapor:2017gdk,Dapor:2017rwv,zhang2019bouncing,Liegener:2019jhj,Han:2019feb,Dapor:2020jvc,Han:2020iwk,Liegener:2020dcg}. In particular, the quantum dynamics of the reduced phase space LQG is formulated on the cubic lattice $\gamma$ as a path integral \cite{han2020effective,Han:2020chr}
\begin{equation}
A_{[g],\left[g^{\prime}\right]}=\int \mathrm{d} h  [\mathrm{d} g]\, \nu[g]\, e^{S[g, h] / \ell_{\rm p}^2},\label{Agg0}
\end{equation}
which is the canonical-LQG analog of the spinfoam formulation. $A_{[g],\left[g^{\prime}\right]}$ is the transition amplitude of $\widehat{\bf H}$ between the initial and final SU(2) gauge invariant coherent states $|[g]\rangle,|\left[g^{\prime}\right]\rangle$. The integration variables contains trajectories of $g\in\mathscr{P}_{red,\gamma}$ and SU(2) gauge transformations $h$ on $\gamma$. $\nu[g]$ is a measure factor. As a feature of the path integral formula \eqref{Agg0}, the SU(2) gauge invariant amplitude $A_{[g],\left[g^{\prime}\right]}$ is expressed as the integral in terms of SU(2) gauge non-invariant variables $g$ and $h$. The action $S[g,h]$ is linear to the expectation value $\langle \widehat{\bf H} \rangle$ at SU(2) gauge non-invariant coherent states when trajectories of $g$ are continuous in time. In contrast to usual path integrals of quantum field theories, $S[g,h]$ contains  {not only the 1-loop quantum correction but also the $O(\hbar)$ correction from $\langle \widehat{\bf H} \rangle$.} Our work precisely computes this $O(\hbar)$ correction from $\langle \widehat{\bf H} \rangle$ in $S[g,h]$  in the case of cosmological dynamics.  

The semiclassical limit $\hbar\to0$ of $A_{[g],\left[g^{\prime}\right]}$ gives the equation of motion studied in general in \cite{Han:2020chr} and applied in cosmology in \cite{han2020effective,Han:2020iwk}. The cosmological dynamics as $\hbar\to0$ gives the $\mu_0$-scheme effective cosmological dynamics which reduces to the classical FRLW cosmology at low energy density. As the next step, it is important to discover the $O(\hbar)$ correction to the effective cosmological dynamics. The effective dynamics with $O(\hbar)$ correction can be obtained by the \emph{quantum effective action} $\Gamma$ \cite{peskin1995introduction} from the path integral \eqref{Agg0}. Perturbatively, the $O(\hbar)$ in $\Gamma$ for cosmology contains 3 contributions: (1) $O(\hbar)$ in $S[g,h]$ which is computed in this work, (2) $\log \nu[g]$ where $\nu[g]$ has been given explicitly in \cite{han2020effective}, and (3) $\frac{1}{2}\log\det(\mathfrak{H})$ where the ``1-loop determinant'' $\det(\mathfrak{H})$ is the determinant of the Hessian matrix $\mathfrak{H}$ of $S[g,h]$. The $g$-$g$ matrix elements in $\mathfrak{H}$ is computed in \cite{Han:2020iwk}. The study of $\frac{1}{2}\log\det(\mathfrak{H})$ is postponed to the future work. Therefore from the perspective of the quantum correction in the effective cosmological dynamics, our work computes an important part in the $O(\hbar)$ of the quantum effective action $\Gamma$.    

After introducing several motivations in the above, let us summarize the key steps in our computation: First of all, an important complication in $\widehat{H[N]}$ is the volume operator $\hat{V}_v=\sqrt{|\hat{Q}_v|}$ which contains the square-root and absolute-value so that $\widehat{H[N]}$ is non-polynomial. In the case of the coherent state expectation value $\langle\widehat{H[N]}\rangle$, this issue is overcome by substituting $\hat{V}_v$ with the semiclassical expansion \cite{giesel2007algebraic}
\begin{equation}\label{eq:VGT}
\begin{aligned}
\hat V_{GT}^{(v)}=&\langle \hat Q_v\rangle^{2q}\left[1+\sum_{n=1}^{2k+1}(-1)^{n+1}\frac{q(1-q)\cdots(n-1-q)}{n!}\left(\frac{\hat Q_v^2}{\langle \hat Q_v\rangle^2}-1\right)^n\right]+O(\hbar^{k+1})
\end{aligned}
\end{equation}
where $\hat{Q}_v$ is a polynomial of flux operators and $q=1/4$. Inserting in $\hat V_{GT}^{(v)}$ truncated at a finite $k$ reduces $\langle\widehat{H[N]}\rangle$ to the expectation value of a polynomial operator.  {Note that the current work uses the volume operator introduced in \cite{ashtekar1997quantumII}, which is the same as that studied in \cite{giesel2007algebraic}. However, this volume operator is not the unique  choice to define $\widehat{H[N]}$. For instance, one can use the volume operators given in  \cite{rovelli1995discreteness,yang2016new}. Even though it is  still open that whether these operators can be replaced by some operator polynomials as far as the expectation value is concerned, it is expected that these operators should give the same leading order result in $\hbar$ if they posses desirable semiclassial properties.
}

The resulting polynomial sums a huge number of terms ($\sim 10^{19}$), each of which is a monomial of holonomies and fluxes. Computing expectation values of all individual terms would give a large computational complexity. The complexity mainly comes from the Lorentzian part of $\widehat{H[N]}$. Several key methods in our work are used to reduce the number of terms needed for computation:

\begin{itemize}

\item The expectation value of every monomial term in $\widehat{H[N]}$ can be factorized into expectation values of holonomy-flux monomials on individual edges. We only need to compute expectation values of various types of monomials on a single edge. We further reduce the number of types by the commutation relations, and derive several general formulae for the expectation values of resulting types (see Section \ref{Expectation values of operators on one edge}).

\item We develop a power-counting argument to find the power of $\hbar$ as the leading order behavior of each expectation value of the monomial operator (see Section \ref{sec:leadingorder}). Since we focus on expanding $\langle\widehat{H[N]}\rangle$ to the linear order in $\hbar$, a large number of monomials only contributing to higher order can be neglected. 

\item When we focus on the expectation values at coherent states peaked at homogeneous and isotropic data. There are a large amount of symmetries which identify different terms and significantly reduce the number of terms needed for computation (see Secton \ref{sec:cosmodel}).   

\end{itemize}

Our method exponentially reduces the computational complexity and is particularly useful in computing the expectation value of Lorentzian part $\widehat{H_L[N]}$ in $\widehat{H[N]}$. In Section \ref{Introducing the algorithm: A case study}, we demonstrate the reduction in a typical example which is an operator contains $3^{3m-1}$ ($m$ can be large) monomials. By applying our method, only 5 monomials are finally needed for computing the expectation values.  

The purpose of this paper is to give an overview our computation and present the results. The detailed derivations are presented in \cite{toappear}. Our computation is carried out in Mathematica on the High Performance Computation server with two 48-Core Processors (AMD EPYC 7642). The Mathematica codes are available at \cite{github}.

The resulting $O(\hbar)$ quantum correction in $\langle\widehat{H[N]}\rangle$ is summarized in Section \ref{sec:results}. In order to demonstrate the physical application of our results and effects from the $O(\hbar)$ correction, we adopt the proposal in \cite{Dapor:2017rwv}: We view $\Re\langle\widehat{H[1]}\rangle$ in \eqref{hamexp0} as the effective Hamiltonian function on the 2-dimensional phase space $\mathscr{P}_{cos}$ of homogeneous and isotropic cosmology. $\Re\langle\widehat{H[1]}\rangle$ generates the Hamiltonian time evolution on $\mathscr{P}_{cos}$. We plot the time evolution of the homogeneous spatial volume, and compare with the evolution generated by $\langle\widehat{H[1]}\rangle$ at $\hbar\to0$. The comparison demonstrates the effect from the $O(\hbar)$ correction in $\langle\widehat{H[1]}\rangle$ (see Section \ref{sec:results} for details). We emphasize that here the proposal that we adopt for the cosmological evolution is not as rigorous as the path integral formula \eqref{Agg0}, and we have argued above that the $O(\hbar)$ correction in $\langle\widehat{H[1]}\rangle$ is only a partial contribution to the quantum effective action $\Gamma$  which ultimately determines the quantum effect in the dynamics. The cosmological dynamics studied in Section \ref{sec:results} is only for the purpose of displaying the effect of the $O(\hbar)$ correction in $\langle\widehat{H[1]}\rangle$, and is not yet a rigorous prediction from the principle of LQG.

The paper is organized as follows. Secion \ref{sec:one} reviews briefly the theory of LQG on a cubic lattice, including the Hamiltonian and the coherent state. In Section \ref{Expectation values of operators on one edge}, we describe the procedure to compute the expectation value of operators at a single edge. In Secion \ref{sec:leadingorder}, we develop a power-counting argument to reduce the computational complexity. Section \ref{sec:cosmodel} discusses $\langle\widehat{H[1]}\rangle$ at coherent states peaked at homogeneous and isotropic data, and discuss the symmetries which further reduce the computational complexity. Section \ref{sec:results} summarizes the results of the quantum correction in $\langle\widehat{H[1]}\rangle$. In Sec. \ref{sec:conclusion}, we conclude and discuss a few future perspectives. \\

Note added in proof: During finalizing this paper, we became aware that the $O(\hbar)$ correction of the Euclidean part $\langle\widehat{H_E}\rangle$ in $\langle\widehat{H[1]}\rangle$ was derived independently in \cite{Liegener:2020dcg} 
\footnote{The result of $\langle\widehat{H_E}\rangle$ in \cite{Liegener:2020dcg} is the same as ours (see \eqref{euclidhamres}) up to an overall constant and $(\eta,\xi)\to(-\eta, -\xi)$.}.

\section{Preliminaries}\label{sec:one}

\subsection{Quantization and Hamiltonian}

Classically general relativity can be reformulated with the Ashtekar-Barbero variables $(A_a^i,E^a_i)$ consisting of an SU$(2)$ connection $A^i_a$ and its canonically conjugate densitized triad field $E_i^a$ defined on the spatial manifold $\Sigma$ \cite{barbero1995real}. 
We choose the coordinate denoted by $(x,y,z)$ on $\Sigma$.  Let $\gamma\subset \Sigma$ be a finite cubic lattice whose edges are parallel to the axes of the coordinates. The sets of edges and vertices in $\gamma$ are denoted by $E(\gamma)$ and $V(\gamma)$. Taking advantage of $\gamma$, we define holonomies along the edges of $\gamma$, 
\begin{equation}\label{eq:holonomy}
h_e(A)=\mathcal P\exp\int_e A=1+\sum_{n=1}^\infty \int_0^1\dd t_n\int_0^{t_n}\dd t_{n-1}\cdots\int_0^{t_2}\dd t_1 A(t_1)\cdots A(t_n),\ \forall e\in E(\gamma),
\end{equation}
and gauge covariant fluxes on the 2-faces $S_e$ in the dual lattices $\gamma^*$,
\begin{equation}\label{eq:flux}
\begin{aligned}
p^i_\s(e):=&-\frac{2}{\beta a^2}\tr\left[\tau^i\int_{S_e} \dd x^a\dd x^b\varepsilon_{abc} h(\rho^\s_e(\sigma) ) E^c(\sigma) h(\rho^\s_e(\sigma)^{-1})\right],
\end{aligned}
\end{equation}
where $S_e\in\gamma^*$ is the 2-face, $\rho^\s(\sigma): [0,1]\to \Sigma$ is a path connecting  the source point $\s_e\in e$ to $\sigma\in S_e$ such that $\rho_e^\s(\sigma):[0,1/2]\to e$ and $\rho_e^\s(\sigma):[1/2,1]\to S_e$.  {The parameter $a$,
identified with a length scale of the system, is chosen to be (cosmological constant)$^{-1/2}$ in our paper. 
}
Alternatively, one can choose the target point $\t_e\in e$ rather than $s_e$ to define
\begin{equation}
\begin{aligned}
p^i_\t(e):=&\frac{2}{\beta a^2}\tr\left[\tau^i\int_{S_e} \dd x^a\dd x^b\varepsilon_{abc} h(\rho^\t_e(\sigma) ) E^c(\sigma) h(\rho^\t_e(\sigma)^{-1})\right].
\end{aligned}
\end{equation}
where $\rho^\t(\sigma): [0,1]\to \Sigma$ is a path connecting  the target $\t_e\in e$ to $\sigma\in S_e$ such that $\rho_e^\t(\sigma):[0,1/2]\to e$ and $\rho_e^\t(\sigma):[1/2,1]\to S_e$. 
%

The quantization of this classical lattice theory gives us LQG based on the graph $\gamma$. The Hilbert space $\mathcal H_\gamma$ consists of the square integrable functions of the holonomies. Given two functions $\psi_i: \{h_e\}_{e\in E(\gamma)}\to \mathbb C$, the inner produce is
\begin{equation}
\langle \psi_1|\psi_2\rangle=\int_{\mathrm{SU}(2)^{|E(\gamma)|}} \dd\muh\,\overline{\psi_1(\{h_e\}_{e\in E(\gamma)})}\psi_2(\{h_e\}_{e\in E(\gamma)})
\end{equation}
where $|E(\gamma)|$ denote the number of elements (i.e. cardinality) of $E(\gamma)$ and $\muh$ is the Haar measure. $\mathcal H_\gamma$ is the kinematical Hilbert space in the operator-constraint formalism of the canonical LQG. However $\mathcal H_\gamma$ modulo gauge transformations is the physical Hilbert space in the reduced phase space LQG, where $h_e(A)$, $p_{\s}^i(e)$ and $p_\t^i(e)$ are Dirac observables from the deparametrization by coupling to clock fields \cite{giesel2010algebraic}.

On $\mathcal H_\gamma$, the holonomy becomes the multiplication operator and, $p_\s^i(e)$ and $p_\t^i(e)$ are quantized as the right- and left-invariant vector field, namely
\begin{equation}\label{eq:ptps}
\begin{aligned}
(\hat p_\s^i(e)\psi)(h_{e'},\cdots,h_e,\cdots,h_{e''})&=i t\left.\frac{\dd}{\dd\epsilon}\right|_{\epsilon=0}\psi(h_{e'},\cdots,e^{\epsilon\tau^i}h_e,\cdots,h_{e''})\\
(\hat p_\t^i(e)\psi)(h_{e'},\cdots,h_e,\cdots,h_{e''})&=-i t\left.\frac{\dd}{\dd\epsilon}\right|_{\epsilon=0}\psi(h_{e'},\cdots,h_ee^{\epsilon\tau^i},\cdots,h_{e''})
\end{aligned}
\end{equation}
where $t=\kappa\hbar/a^2=:\Pl^2/a^2$ 
and $\tau^j=(-i/2)\sigma^j$ with $\sigma^j$ the Pauli matrix. 
The commutators between the basic operators are
\begin{equation}\label{eq:commutators}
\begin{aligned}
[h(e),h(e')]&=0=[\hat p_\s^i(e),p_\t^j(e')]\\
[\hat p_\s^i(e),\hat p_\s^j(e')]&=-it\delta_{ee'}\epsilon_{ijk}\hat p_\s^k(e),\\
[\hat p_\t^i(e),\hat p_\t^j(e')]&=-it\delta_{ee'}\epsilon_{ijk}\hat p_\t^k(e),\\
[\hat p_\s^i(e),h(e')]&=it\delta_{ee'}\tau^ih(e),\\
[\hat p_\t^i(e),h(e')]&=-it\delta_{ee'}h(e)\tau^i.
\end{aligned}
\end{equation}
It is useful to introduce the flux operators with respect to the spherical basis. We define
\begin{equation}\label{eq:sphericalbasis}
\hat p_v^{\pm 1}(e):=\mp\frac{1}{\sqrt{2}}\left(\hat p_v^x(e)\pm i \hat p_v^y(e)\right),\ \hat p_v^0(e)=\hat p_v^z(e)
\end{equation}
with $v=\s,\t$. In the following context, $\alpha,\beta,\cdots=0,\pm 1$ is used to denote the indices in the spherical basis, and $i,j,k\cdots=1,2,3$, the indices in the Cartesian basis.

In the operator-constraint formalism, the dynamics of LQG is encoded in the Hamiltonian constraint, which can be written as 
\begin{equation}
\widehat{H[N]}=\widehat{H_E[N]}+(1+\beta^2)\widehat{H_L[N]}
\end{equation}
where $\widehat{H_E[N]}$ is called the Euclidean part and $\widehat{H_L[N]}$ is the Lorentzian part. $N$ is the smeared function. $\widehat{H[N]}$ is constructed by using the Thiemann's trick  \cite{thiemann1998quantum,Giesel:2006uj}. The operator corresponding to the Euclidean part is
\begin{equation}
\begin{aligned}
&\widehat{H_E[N]}=\frac{1}{2i\beta a^2 t}\sum_{v\in V(\gamma)} N(v)\sum_{e_I,e_J,e_K \text{ at } v}\epsilon^{IJK}\tr(h_{\alpha_{IJ}}h_{e_K}[\hat V_v,h_{e_K}^{-1}])
\end{aligned}
\end{equation}
where  $e_I$, $e_J$ and $e_K$ are oriented to be outgoing from $v$, $\epsilon^{IJK}=\sgn[\det (e_I\wedge e_J\wedge e_K)]$, $\alpha_{IJ}$ is the minimal loop around a plaquette consisting of $e_I$ and $e_J$, where it goes out via $e_I$ and comes back through $e_J$, taking $v$ as its end point. The volume operator $\hat V_v$ at $v$ reads
\begin{equation}
\hat V_v=\sqrt{|\hat Q_v|}
\end{equation}
where the operator $\hat Q_v$ in terms of the flux operators with respect to the spherical basis \eqref{eq:sphericalbasis} is
\begin{equation}\label{eq:qdefinition}
\begin{aligned}
&\hat Q_v=-i(\beta a^2)^3\varepsilon_{\alpha\beta\gamma} \frac{\hat p_\s^\alpha(e_x^+)-\hat p_\t^\alpha(e_x^-)}{2}\frac{\hat p_\s^\beta(e_y^+)-\hat p_\s^\beta(e_y^-)}{2}\frac{\hat p_\s^\gamma(e_z^+)-\hat p_\s^\gamma(e_z^-)}{2}
\end{aligned}
\end{equation}
with $\varepsilon_{\alpha\beta\gamma}$ defined by $\varepsilon_{-1,0,1}=1$.
With the same notion, the Lorentzian part reads
\begin{equation}
\begin{aligned}
\widehat{H_L[N]}=\frac{-2}{i\beta^7 a^{10} t^5}\sum_v N(v)\sum_{e_I,e_J,e_K \text{ at } v}\varepsilon^{IJK}\tr( [h_{e_I},[\hat V,\hat H_E]]h_{e_I}^{-1} [h_{e_J},[\hat V,\hat H_E]]h_{e_J}^{-1}[h_{e_K},\hat V_v]h_{e_K}^{-1}).
\end{aligned}
\end{equation}

%
%
%

\subsection{Coherent states}\label{sec:two}

Choosing a canonical orientation for each edge $e\in E(\gamma)$, the classical phase space based on the graph $\gamma$ is 
\begin{equation}
\Gamma_\gamma\cong [{\rm SL}(2,\mathbb C)]^{ |E(\gamma)|}.
\end{equation}
The complexifier coherent state $\Psi_\g$ is \cite{thiemann2001gauge} 
\begin{equation}\label{eq:coherentstate}
\Psi_\g=\bigotimes_{e\in E(\gamma)}\psi^{ t}_{g_e},\qquad \psi^t_{g_e}(h_e)=\sum_j d_je^{-\frac{t}{2}j(j+1)}\chi_j(g_e h_e^{-1})
\end{equation}
where $\psi^{ t}_{g_e}$ is the SU(2) coherent state at the edge $e$. The character $\chi_j(g_eh_e^{-1})$ is the trace of the $j$-representation of $g_eh_e^{-1}$. The property $\chi_j(g_e h_e^{-1})=\chi_j(g_e^{-1}h_e)$ leads to the useful relation $$\psi_{g_e}(h_e)=\psi_{g_{e^{-1}}}(h_{e^{-1}}).$$

 {Given $g=e^{ip_k\tau^k}u\in$SL($2,\mathbb C$),
one can find $n^s\in$SU(2)  such that $g=n^\s e^{i\eta\tau_3}(n^\s)^{-1}u$. Due to $(n^\s)^{-1}u\in $SU(2), one can decompose it as $(n^\s)^{-1}u=e^{-\xi\tau_3} (n^\t)^{-1}$. Finally, we have  
\begin{equation}\label{eq:decomposition}
g=n^\s e^{i(\eta+i\xi)\tau_3}(n^\t)^{-1}.
\end{equation}
In this decomposition, one has $n^\s\tau_3 (n^\s)^{-1}=\vec p\cdot\vec \tau/\eta$, which implies $|\eta|=\sqrt{\vec p\cdot\vec p}$. In the current we choose the convention with $\eta=-\sqrt{\vec p\cdot\vec p}$. Moreover, this decomposition associates to  $n^v\in$SU(2) ($v=\s,\t$) a vector $\vec n^v$:
\begin{equation}
n^v\tau_3 (n^v)^{-1}=\vec n^v\cdot \vec\tau,\ v=\s,\t.
\end{equation}
Indeed, one has $\vec n^\s=\vec p/\eta$ and $\vec n^\t=(u^{-1}\triangleright \vec p)/\eta$ with $u^{-1}\triangleright \vec p$ denoting the vector of $\vec p$ rotated by $u^{-1}$.
}
It is shown in \cite{thiemann2001gaugeIII} and is revisited shortly that
\begin{equation}
\langle \psi_{g_e}^t|\vec{\hat p}_\s(e)|\psi^t_{g_e}\rangle=-\eta_e\vec n_e^\s+O(t),\ \langle \psi_{g_e}^t|\vec{\hat p}_\t(e)|\psi_{g_e}^t\rangle= \eta_e\vec n_e^\t+O(t)
\end{equation}
which indicates that $\eta_e\vec n_e^\t$ is the classical limit of the flux operator at $e$.
 
The following properties of the $\psi_g^t$  \cite{thiemann2001gauge,thiemann2001gaugeII,thiemann2001gaugeIII} are useful in our analysis. Firstly, the inner product of these states read
\begin{equation}\label{eq:innerproduct}
\langle\psi_{g_1}^t|\psi_{g_2}^t\rangle=\psi^{2t}_{g_1^\dagger g_2}(1)= \frac{2\sqrt{\pi } e^{t/4}}{t^{3/2}}\, \frac{\zeta\,e^{\frac{\zeta^2}{t}}}{\sinh(\zeta)}+O(t^\infty)
\end{equation}
where $\tr(g_1^\dagger g_2)=2\cosh(\zeta)$ and $\Im(\zeta)\in[0,\pi]$ with $\Im(\zeta)$ the imaginary part of $\zeta$ \footnote{Here we used the following result shown in \cite{thiemann2001gaugeIII}. For any complex number $z=R+iI$, there exist real numbers $s\in\mathbb R$ and $\phi\in[0,\pi]$  such that  $\cosh(s+i\phi)=z$. $s$ and $\phi$ are uniquely determined except in the case $I=0$ and $|R|>1$ in which case the $s$ is determined up to its sign.}. Consequently, the norm of the coherent state is
\begin{equation}\label{eq:norm}
\begin{aligned}
\langle 1\rangle_{g}:=\langle\psi^t_g|\psi^t_g\rangle= \frac{2\sqrt{\pi } e^{ t/4}}{t^{3/2}}\frac{ p e^{\frac{ p ^2}{t}}}{\sinh(p)}+O(t^\infty),
\end{aligned}
\end{equation}
where $p=\sqrt{\vec{p}\cdot\vec{p}}$. Secondly,
$\psi_g^t$ satisfy the completeness condition
\begin{equation}\label{eq:completerelation}
\int \dd\nu_t(g)|\psi^t_g\rangle\langle\psi^t_g|=\mathbb I,
\end{equation}
where the measure $\dd\nu_t(g)$ is  
\begin{equation}\label{eq:measure}
\dd\nu_t(g)=\frac{2\sqrt{2} e^{-t/4}}{(2\pi t)^{3/2}}\frac{\sinh(p)}{p}e^{-\frac{p^2}{t}}\dd\mu_H(u)\dd ^3p=\frac{2}{\langle 1\rangle_g\pi t^3}\dd\mu_H(u)\dd ^3p.
\end{equation}

Let us complete this section with some discussions on the volume operator contained in the Hamiltonian operator $\widehat{H[N]}$.  
Because of the square root in the definition of the volume operator, matrix elements of these operators are difficult to compute analytically. However, as far as the coherent state expectation value is concerned, the volume operators $\hat{V}_v$ in $\widehat{H[N]}$ can be replaced by Giesel-Thiemann's volume $\hat V_{GT}^{(v)}$  {as aforementioned in \eqref{eq:VGT}} with $q=1/4$.
Substituting $\hat{V}_v$ with $\hat V_{GT}^{(v)}$ truncated at finite $k$ transforms $\widehat{H[N]}$ to be a polynomial of holonomies and fluxes. Up to higher order in $t$, it is now manageable to compute the expectation value of $\widehat{H[N]}$, through computing the expectation value of a polynomial of holonomies and fluxes.

\section{Expectation values of operators on one edge}\label{Expectation values of operators on one edge}

 {Because we only consider gauge \textit{variant} coherent state, then the gauge \textit{variant} coherent state can be written simply as the tensor product of SU(2) coherent state on each edge. As a consequence, computing the expectation value of $\widehat{H[N]}$ with respect to a coherent state can be reduced to computing expectation values of operator monomials on individual edges.  In this section, let us firstly focus on the expectation value of operators on one edge. }

Given a monomial of holonomies and fluxes on an edge $e$, its expectation value with respect to the coherent state $\psi^t_{g_e}$ labelled by $g_e=n_e^\s e^{i z_e\tau_3}(n_e^\t)^{-1}$ relates to its expectation value with respect to $\psi^t_{e^{iz_e\tau_3}}\equiv \psi^t_{z_e}$, by a gauge transformation generated by $n_e^\s$ and $n_e^\t$ \cite{Dapor:2017gdk}:
\begin{equation}\label{eq:basicformula}
\begin{aligned}
&\langle \psi^t_{g_e}|P(\{\hat p_\s^{\alpha_i}(e)\},\{\hat p_\t^{\alpha_j}(e)\},\{D^{\iota_k}_{a_k b_k}(h_e)\})|\psi^t_{g_e}\rangle\\
=&\langle \psi^t_{z_e}|P(\{\hat p_\s^{\beta_i}(e)D^1_{\beta_i\alpha_i}((n_e^\s)^{-1})\},\{\hat p_\t^{\beta_j}(e)D^1_{\beta_j\alpha_j}((n_e^\t)^{-1})\},\{D^{\iota_k}_{a_k c_k}(n_e^\s)D^{\iota_k}_{c_k d_k}(h_e)D^{\iota_k}_{d_k b_k}((n_e^\t)^{-1})\})|\psi^t_{z_e}\rangle,
\end{aligned}
\end{equation}
where $P(x,y,z)$ represents any monomial of $x$, $y$ and $z$. In the following context, we denote 
\begin{equation}
\langle \psi^t_{z_e}|\hat F_e|\psi^t_{z_e}\rangle=:\langle \hat F_e\rangle_{z_e}. 
\end{equation}
According to \eqref{eq:basicformula}, 
in order to compute $\langle \psi^t_{g_e}|\hat F_e|\psi^t_{g_e}\rangle$,
it is sufficient to compute expectation values with respect to $\psi^t_{z_e}$. The algorithm to compute the expectation value with respect to $\psi^t_{z_e}$ is divided into three steps. 
\subsection{the first step}
Let $\hat O$ be a monomial of flux and holonomy operators. To compute the expectation value of $\hat O$,
we firstly move all the holonomies operators to the right, re-ordering the operators, by implementing the basic commutation relations  \eqref{eq:commutators}. 
To this end, we need 
\begin{equation}\label{eq:changeO}
\begin{aligned}
\hat O=\hat O_2\cdots\hat O_m\hat O_1+\sum_{k=1}^{m-1}\sum_{\mathcal I_k} \left(\prod_{l \in\mathcal I-\mathcal I_k}\hat O_l\right)[[\cdots[[\hat O_1,\hat O_{i_1}],\hat O_{i_2}]\cdots],\hat O_{i_k}].
\end{aligned}
\end{equation}
with $\hat O=\hat O_1\hat O_2\cdots\hat O_m$, $\mathcal I:=\{2,\cdots,m\}$ and $\mathcal I_k:=\{i_1,i_2,\cdots,i_k\}\subset \mathcal I$ containing $k$ elements with $i_1<i_2<\cdots<i_k$. 
In Eq. \eqref{eq:changeO}, the terms at $k$ carry $k$-fold commutator. Due to the factor $t$ in the right hand side of the commutation relations \eqref{eq:commutators},
the $k$-fold commutator produces a factor $t^k$ in the final result, which implies that the contributions of these terms to the expectation value of $\hat O$ are at least at $t^k$-order. 
\subsection{the second step}
Once all of the holonomies are moved to the right, they can be merged into a single Winger-D matrix by using the angular momentum recoupling theory 
\begin{equation}\label{eq:coupling}
\begin{aligned}
D^{j_1}_{m_1n_1}(h)D^{j_2}_{m_2n_2}(h)=\sum_{J={j_1-j_2}}^{j_1+j_2}d_J(-1)^{M-N}\left(
\begin{array}{ccc}
j_1&j_2&J\\
m_1&m_2&-M
\end{array}
\right)\left(
\begin{array}{ccc}
j_1&j_2&J\\
n_1&n_2&-N
\end{array}
\right)D^J_{MN}(h).
\end{aligned}
\end{equation}
Thus it is sufficient to consider the operators of the form $\hat O_1=\hat p_\s^{\alpha_1}(e)\cdots \hat p_\s^{\alpha_m}(e)\hat p_\t^{\beta_1}(e)\cdots \hat p_\t^{\beta_n}(e)D^{\iota}_{ab}(h_e)$. 
To calculate the expectation value of $\hat O_1$, we transform $\hat p_\t^\alpha(e)$ to $\hat{p}_\s^\alpha(e)$ with the identity \cite{toappear}
\begin{equation}\label{eq:pspth}
\begin{aligned}
&\langle p_\s^{\alpha_1}(e)\cdots p_\s^{\alpha_m}(e)p_\t^{\beta_1}(e)\cdots p_\t^{\beta_n}(e)D^{\iota}_{ab}(h_e)\rangle_{z_e}\\
=&(-1)^n e^{-(\beta_1+\cdots+\beta_n)\overline{z_e}} \langle p_\s^{\beta_n}(e)\cdots p_\s^{\beta_1}(e) p_\s^{\alpha_1}(e)\cdots p_\s^{\alpha_m}(e)D^{\iota}_{ab}(h_e)\rangle_{z_e}
\end{aligned}
\end{equation}
where the factor $(-1)^n$ in the right hand side is because of the minus sign in the definition of $p_\t^\alpha(e)$. In summary, the second step is to apply Eqs. \eqref{eq:coupling} and \eqref{eq:pspth} to simplify the concerning operator into a linear combination of operators taking the form 
$$\hat F^{\alpha_1\cdots \alpha_m}_{\iota a b}=\hat p_\s^{\alpha_1}(e)\cdots \hat p_\s^{\alpha_m}(e)D^{\iota}_{ab}(h_e).$$
The the last step is to calculate the expectation value of $\hat F^{\alpha_1\cdots \alpha_m}_{\iota a b}$.

\subsection{the third step}
The strategy to compute the expectation value of the operator $\hat F^{\alpha_1\cdots \alpha_m}_{\iota a b}$ is to write the action of the operator on the coherent state $\psi_{z_e}$ in terms of Wigner $3jm$-symbols \cite{yang2017graphical,toappear}, 
\begin{equation}\label{eq:F0}
\langle \hat F^{\alpha_1\cdots\alpha_m}_{\iota a b}\rangle_{z_e}=t^me^{bz_e}\sum_{j,j'}e^{-\frac{t}{2}(j(j+1)+j'(j'+1))}F_\iota(j,j',\frac{\partial_\eta}{2})\frac{\sinh((2j'+1)\eta)}{\sinh(\eta)}
\end{equation}
with $\hat F_{\iota}$ expressed by Wigner $2j$- and $3j$- symbols and $j$, for each fixed half integer $j'$, taking values of $|j'-\iota|,|j'-\iota|+1,\cdots,|j'+\iota|$. The summation over $j'$ will be computed with the Poisson summation formula. To do this, we extend the sum from $j'\geq 0$ to $j'\leq 0$ by applying the symmetry  
\begin{equation}
F_{\iota}(j'+\Delta,j',k')=-F_{\iota}(j'-\Delta,j',k')\Big|_{j'\to -j'-1},
\end{equation}
which is proved in details in \cite{toappear}. 
Then, summing over $j'$ with $j=j'+\Delta$ and $j=j'-\Delta$ can be combined and rewritten as summing over all positive and negative half-integers $j'$, so that the Poisson summation formula can be applied. 
One can refer to \cite{toappear} for the explicit expression of $F_\iota$.

After applying the Poisson summation formula, Eq. \eqref{eq:F0} will simplified to take the form:
\begin{equation}\label{eq:expectedgeneral}
\langle \hat F^{\alpha_1\cdots\alpha_m}_{\iota a b}\rangle_{z_e}=\sum_{\substack{0\leq d\leq \iota\\ d+\iota\in \mathbb Z}}\int\dd x W_d(x)e^{-\frac{1}{4} t \left(x^2-2 d x\right)}F_\iota(\frac{x-1}{2}-d,\frac{x-1}{2},\frac{\partial_\eta}{2})\frac{\sinh(x\eta)}{\sinh(\eta)}+O(t^{-\infty})
\end{equation}
with some function $W_d$. Then, according to the explicit result in \cite{toappear}, for at least $\iota\leq 20$, $F_\iota(\frac{x-1}{2}-d,\frac{x-1}{2},\frac{\partial_\eta}{2})\frac{\sinh(x\eta)}{\sinh(\eta)}$, as a function of $x$,  is a summation of polynomials of $x$ and and the functions $\frac{\sinh(\eta (x-n))}{x-n}$ with $n$ being some integers. Thus, the integral in Eq. \eqref{eq:expectedgeneral} is a linear combination of integrals 
$I_1=\int_{-\infty}^\infty \dd x\, e^{-a x^2+bx } \frac{\sinh(x \eta)}{x}$
and
$I_2=\int_{-\infty}^\infty\dd x\, e^{- a x^2+b x} x^n e^{\pm \eta x}$
where $a>0$, $b\in\mathbb R$. The integrals $I_1$ and $I_2$ are easily to be computed. The only remaining problem is how to realize the concrete linear combination form of $I_1$ and $I_2$, which can be quite technical and illustrated by the derivation of $\langle\hat F^{\alpha_1\cdots\alpha_m}_{\iota a b}\rangle_{z_e}$ for $\iota=1/2$ and $\iota =1$ in \cite{toappear}.

 {We would like to compare our algorithm presented here with the known results in \cite{Liegener:2019jhj}.  At first, our algorithm generalizes the known results in literature \cite{Liegener:2019jhj}, in the sense that our formula gives the results for arbitrary lists $\{\alpha_i\}_{i=1}^m$ of flux indices and triples $(\iota,a,b)$ with at least $\iota\leq 20$. Moreover, with our formula, one can get the expectation values to arbitrary order of $t$. However, in  \cite{Liegener:2019jhj}, the authors give only the results for the special cases where the list $\{\alpha_i\}_{i=1}^m$ contains at most either a single $-1$, or a single $1$, or a pair of $(-1,1)$. They are all the cases such that  the expectation values have non-vanishing $O(t^0)$ or $O(t)$-term. Other cases are also interesting when we study the higher-order correction, even though the higher-order  correction  is beyond the present work. Our formula reduce to these known results at the special cases. The current work only use these special cases, but our codes \cite{github} are designed based on the generalization algorithm, since the generalized formulae have the potential for computing higher-order correction.}

In summary, we have outlined a general procedure to compute the expectation value of any monomial of holonomies and fluxes. According to this mechanism, we develop an algorithm in Mathematica to obtain analytical result  \cite{github}.

\section{Power counting}\label{sec:leadingorder}

Applying $V_{GT}^{(v)} $ in \eqref{eq:VGT} and truncating to a finite $k$, $\widehat{H[N]}$ is of the form $\sum_{\vec{\alpha}}\mathcal T^{\alpha_1\cdots\alpha_m}\hat O_{\alpha_1\cdots \alpha_m}$ with $\mathcal T^{\alpha_1\cdots\alpha_m}$ being some numerical factors and $\hat O_{\alpha_1\cdots \alpha_m}$ being some polynomial operators of holonomies and fluxes.  {We consider gauge \textit{variant} coherent state, namely, we do not introduce  intertwiners and group averaging procedure. Then, the gauge \textit{variant} coherent state can be written simply as the tensor product of SU(2) coherent state on each edge, so that
the expectation can be factored into single edge expectation values. Thus the algorithm introduced in Sec. \ref{Expectation values of operators on one edge} can be applied to calculate the expectation value of  $\hat O_{\alpha_1\cdots \alpha_m}$. 
}
To do this, in principle, we would need to compute the expectation values of $\hat O_{\alpha_1\cdots \alpha_m}$ for all indices ${\alpha_1\cdots \alpha_m}$. The computational complexity comes from the huge number of terms in the sum over $\vec{\alpha}$. Since we only focus on expanding $\langle\widehat{H[N]}\rangle$ to $O(t)$, the complexity can be reduced by certain power-counting argument: We count the least power of $t$ contains in each $\langle\hat O_{\alpha_1\cdots \alpha_m}\rangle$ before explicit computation, then we omit those terms only contribute to higher order than $O(t)$ in $\langle\widehat{H[N]}\rangle$. It turns out that a large degree of complexity can be reduced in this manner. 

To illustrate how we do power counting before explicit computation, let us introduce some conventions at first. 
In this section, we denote $\Psi_\g$ defined in \eqref{eq:coherentstate} by $|\Psi_{\vec g}\rangle$  with $\vec g=\{g_e\}_{e\in E(\gamma)}$, namely
\begin{equation}\label{eq:coherentstatepsig}
|\Psi_{\vec g}\rangle=\bigotimes_{e\in E(\gamma)}|\psi^t_{g_e} \rangle.
\end{equation}
Similarly, $|\Psi_{\vec g^{(i)}}\rangle$ denotes the coherent state in which $|\psi^t_{g_e^{(i)}}\rangle$ is at edge $e$. Let $\hat O$ take the form of
\begin{equation}\label{eq:operatorO}
\hat O=\hat O_1\hat O_2\cdots\hat O_k,
\end{equation}
with $\hat O_i$ being  polynomial of fluxes and holonimies. Besides, we assume that the matrix elements $\hat O_i$ with respect to coherent states $\psi^t_{g_e^{(1)}}$ and $\psi^t_{g_e^{(2)}}$ on $e$ take the form 
\begin{equation}\label{eq:matrixelement}
\langle \Psi_{\vec g^{(1)}}|\hat O_i|\Psi_{\vec g^{(2)}}\rangle=\langle \Psi_{\vec g^{(1)}}|\Psi_{\vec g^{(2)}}\rangle\left(E_0^{(i)}(\vec g^{(1)},\vec g^{(2)})+t E_1^{(i)}(\vec g^{(1)},\vec g^{(2)})+O(t^\infty)\right).
\end{equation}
Note that this assumption is always satisfied by the operators considered in the current work. Indeed, according to the results in \cite{thiemann2001gaugeIII}, if $\hat O_i$ is just a flux or holonomy on some edge $e$, its matrix elements with respect to coherent states $\psi^t_{g_e^{(1)}}$ and $\psi^t_{g_e^{(2)}}$ on $e$ take the form 
\begin{equation}
\langle\psi^t_{g_e^{(1)}}|\hat O_i|\psi^t_{g_e^{(2)}}\rangle=\langle\psi^t_{g_e^{(1)}}|\psi^t_{g_e^{(2)}}\rangle\left(E_0(g_e^{(1)},g_e^{(2)})+tE_1(g_e^{(1)},g_e^{(2)})+O(t^\infty)\right). 
\end{equation}
Let $N_0$ be the number of operators  $\hat O_m\in \{\hat O_i\}_{i=1}^k$ such that
\begin{equation}\label{eq:operatorom}
\frac{\langle \Psi_{\vec w}|\hat O_m|\Psi_{\vec w}\rangle}{\langle \Psi_{\vec w}|\Psi_{\vec w}\rangle}=O(t),
\end{equation}
where the $O(t^0)$ term vanishes on the RHS.  
Using \eqref{eq:matrixelement}, it turns out tha the expectation value of $\hat O$ with respect to the coherent state $|\Psi_{\vec w}\rangle$ satisfies \cite{toappear}
\begin{equation}\label{thm:leadingordergeneral}
\frac{\langle \Psi_{\vec w}|\hat O|\Psi_{\vec w}\rangle}{\langle \Psi_{\vec w}|\Psi_{\vec w}\rangle}=O(t^n),\ \text{with }n\geq {\floor{\frac{N_0+1}{2}}}
\end{equation}
where $\floor{x}$ is the largest integer no larger than $x$.

In the practical computation of $\langle \widehat{H[N]}\rangle$ with the application of $\hat{V}^{(v)}_{GT}$, these $N_0$ operators $\hat{O}_m$ satisfying \eqref{eq:operatorom} can be the operator $\frac{\hat Q}{\langle\hat Q\rangle}-1$. Then we use Eq. \eqref{thm:leadingordergeneral} to count the power of $t$ for the term with $\left(\frac{\hat Q^2}{\langle\hat Q\rangle^2}-1 \right)^k$ in $\langle \widehat{H[N]}\rangle$. Note that Eq. \eqref{thm:leadingordergeneral} needs the matrix elements of $\hat O_i$. We have to factorize $\frac{\hat Q^2}{\langle\hat Q\rangle^2}-1 =\left(\frac{\hat Q}{\langle\hat Q\rangle}+1 \right)\left(\frac{\hat Q}{\langle\hat Q\rangle}-1 \right)$, because every matrix element of $\hat Q$ is a polynomial of matrix elements of flux operators, while that of $\hat Q^2$ is not.

Moreover, these $N_0$ operators satisfying \eqref{eq:operatorom} can also be $\hat p_\s^{\pm 1}(e)$, $\hat p_\t^{\pm 1}(e)$ and $D^\iota_{ab}(h_e)$ with $a\neq b$ whose expectation values with respect to $\Psi_{\vec \omega}$ vanish. Then this theorem can be used to study the leading order of monomial of holonomies and fluxes.
Let us fix {$\hat p^{\beta}(e)$} to be either $\hat p_\s^\beta(e)$ or $\hat p_\t^\beta(e)$ and, use 
$\mathcal M$ to denote the monomial of holonomies $\{D^{\frac12}_{a_jb_j}(h_{e})\}_{j=1}^k$ and fluxes $\{\hat p^{\beta_i}(e)\}_{i=1}^m$. Let $N_\pm$ be the number of $\hat p^{\pm 1}(e)\in \{\hat p^{\beta_i}(e)\}_{i=1}^m$ and $M_+$ (respectively $M_-$) be the number of $D^{\frac{1}{2}}_{-\frac{1}{2}\frac12}(h_e)\in \{D^{\frac12}_{a_jb_j}(h_{e})\}_{j=1}^k$ (respectively $D^{\frac{1}{2}}_{\frac{1}{2},-\frac12}(h_e)\in \{D^{\frac12}_{a_jb_j}(h_{e})\}_{j=1}^k$). According to our analysis above, the expectation value of $\mathcal M$ with respect to the coherent state $|\psi_{z_e}\rangle$ with $z_e\in \mathbb C$ is non-vanishing provided that 
\begin{equation}
\sum_{i=1}^{m}\beta_i+\sum_{j=1}^k(b_j-a_j)=0. 
\end{equation}
 Thus it has
 \begin{equation}
N_++M_+=N_-+M_-.
\end{equation}
Therefore, this theorem tells us that the leading order the expectation value of $\langle \mathcal M\rangle_{z_e}$ is of order $O(t^{N_++M_+})$ or higher. 

We have more discussions on this case. Since the matrix elements of $\hat p^\alpha(e)$ and $D^{\frac 12}_{ab}(h_e)$ are computable, the results on the leading order of $\mathcal M$ can be calculated more exactly by applying the generalized stationary phase approximation analysis. A straightforward calculation tells us 
\begin{equation}\label{eq:leadingequalmultileading}
 \langle \mathcal M \rangle_{z_e}\cong (\langle \hat p_\s^0(e)\rangle_{z_e})^{N_{0,\s}}(\langle \hat p_\t^0(e)\rangle_{z_e})^{N_{0,\t}}(\langle D^{\frac 12}_{\frac12\frac12}(h_e)\rangle_{z_e})^{M_{0,+}}(\langle D^{\frac 12}_{-\frac12-\frac12}(h_e)\rangle_{z_e})^{M_{0,-}}\langle \mathcal M'\rangle_{z_e}
 \end{equation}  
 where $\cong $ means the leading-order terms, i.e., the $O(t^{M_++N_+})$ terms, of the left and right hand sides are equal to each other, $\mathcal M'$ is the operator resulting from $\mathcal M$ by deleting all factors $\hat p^0(e)$ and $D^{\frac{1}{2}}_{aa}(h_e)$, $N_{0,\s}$ and $N_{0,\t}$ are the number of $\hat p_\s^0(e)$ and $\hat p_\t^0(e)$ in $\mathcal M$, and $M_{0+}$ and $M_{0-}$ the number of $D^{\frac 1 2}_{\frac12 \frac12}(h_e)$ and $D^{\frac 1 2}_{-\frac12 -\frac12}(h_e)$ in $\mathcal M$.

 \section{Cosmological expectation value}\label{sec:cosmodel}
 
We apply our computation of expectation values to coherent states labelled by homogeneous and isotropic data. The symmetry group of the homogeneous and isotropic cosmology is $\mathbb T \rtimes F$ with  $F$ being the isotropy subgroup, $\mathbb T$ being the translation subgroup. Let $S_\gamma$ be the subgroup of $\mathbb T \rtimes F$ preserving $\gamma$. 
A classical state $\g$ is called symmetric with respect to $S_\gamma$ if $s^*\g:=\g \circ s$ are the same as $\g$ up to a gauge transformation for all $s\in S_\gamma$.  By this definition, a classically symmetric state $\g$ takes the form
\begin{equation}\label{eq:coscoh}
\g:e\mapsto g_e=n_e e^{iz\tau_3} n_e^{-1}
\end{equation}
where $n_e\in$SU(2) satisfies
\begin{equation}
n_e\tau_3 n_e^{-1}=\vec n_e\cdot\vec\tau
\end{equation}
 $\vec n_e$ is the unit vector pointing to direction of edge $e$. Then for each $s=(T,f)\in \mathbb T \rtimes F$, it has
 \begin{equation}\label{eq:gaugeAndDiff}
 \g\circ s={\rm Ad}_f\circ \g
 \end{equation}
 where ${\rm Ad}_f \circ \g(e)=f \g(e) f^{-1}$ for all $e\in E(\gamma)$ with $f\in$ SU(2).

\subsection{Symmetries of the expectation value} \label{sec:symmetry1}
Let $\hat F_e$ be an operator of polynomial of fluxes and holonomies on $e$.  For $s=(T,f)\in\mathbb T\rtimes F$, Eq. \eqref{eq:gaugeAndDiff} leads to that 
\begin{equation}\label{eq:DifftoOp}
\langle \psi_{g_{s(e)}}|\hat F_{s(e)}|\psi_{g_{s(e)}}\rangle=\langle \psi_{f g_{e}f^{-1}}|\hat F_{e}|\psi_{fg_{e}f^{-1}}\rangle=f\triangleright \langle \psi_{g_{e}}|\hat F_{e}|\psi_{g_{e}}\rangle
\end{equation}
where $f\triangleright \langle \psi_{g_{e}}|\hat F_{e}|\psi_{g_{e}}\rangle$ means the gauge transformation to the expectation value $\langle \psi_{g_{e}}|\hat F_{e}|\psi_{g_{e}}\rangle$ and Eq. \eqref{eq:basicformula} was used to derive the last equality. 

To expand the expectation value of $\hat H_E$ and $\hat H_L$ to order $O(t)$, we first replace the operator $\hat V_v$ by $\hat V_{GT}^{(v)}$ in \eqref{eq:VGT}. 
Then the Euclidean part $\widehat{H_E[N]}$ can be rewritten in terms of (there is no summation over $I,J,K$ here)
\begin{equation}\label{eq:hen}
\hat{H}_E^{(n)}(v;e_I,e_J,e_K)=\frac{1}{i\beta a^2 t}\epsilon_{IJK}\tr(h_{\alpha_{IJ}}[h_{e_K},\hat Q_v^{2n}]h_{e_K}^{-1}),
\end{equation}
and the Lorentzian part, in terms of 
\begin{equation}\label{eq:hln}
\begin{aligned}
&\hat{H}_L^{(\vec k)}(v;v_1,v_2,v_3,v_4;e_I,e_J,e_K)\\
=&\frac{-2}{i\beta^7 a^{10} t^5}\epsilon^{IJK}\tr( [h_{e_I},[\hat Q_{v_1}^{2k_1},\hat{H}_E^{(k_2)}(v_2)]]h_{e_I}^{-1} [h_{e_J},[\hat Q_{v_3}^{2k_3},\hat{H}_E^{(k_4)}(v_4)]]h_{e_J}^{-1}[h_{e_K},\hat Q_v^{2k_5}]h_{e_K}^{-1})
\end{aligned}
\end{equation}
where $\vec k=(k_1,k_2,k_3,k_4,k_5)$. Let us define 
\begin{equation}\label{eq:HEnv}
\hat H_E^{(n)}(v)=\sum_{e_I,e_J,e_K}\hat H_E^{(n)}(v;e_I,e_J,e_K)
\end{equation}
 and
 \begin{equation}\label{eq:HLnv}
\hat H_L^{(\vec k)}(v)=\sum_{v_1,v_2,v_3,v_4,e_I,e_J,e_K}\hat{H}_L^{(\vec k)}(v;v_1,v_2,v_3,v_4;e_I,e_J,e_K)
 \end{equation}
The Euclidean and Lorentzian parts with the replacement $\hat V_v$ by $\hat V_{GT}^{(v)}$, truncated at a finite $k$ of \eqref{eq:VGT}, are given by linear combinations of $\hat H_E^{(n)}(v)$ and $\hat H_L^{(\vec k)}(v)$ for various $n$ and $\vec k$ respectively.

\subsubsection{Symmetries of the Euclidean part}

By Eq. \eqref{eq:DifftoOp} and the fact that $\hat{H}_E^{(n)}(v;e_I,e_J,e_K)$ is gauge invariant, we obtain that
\begin{equation}\label{eq:HEDiffInv}
\begin{aligned}
\langle \hat{H}_E^{(n)}(v;e_I,e_J,e_K)\rangle&=\langle \hat{H}_E^{(n)}(s(v);s(e_I),s(e_J),s(e_K))\rangle\\
\end{aligned}
\end{equation}
where $\langle\cdot\rangle$ denotes the expectation value with respect to the cosmological coherent state defined by \eqref{eq:coscoh} and $s=(T,f)$ as defined in Eq. \eqref{eq:DifftoOp} is a symmetry of the graph. According to this relation, Eq. \eqref{eq:HEnv} can be simplified as
\begin{equation}\label{eq:HEnvp}
\hat H_E^{(n)}(v)=24(\hat H_E^{(n)}(v;e_x^+,e_y^+,e_z^+)+\hat H_E^{(n)}(v;e_x^+,e_y^+,e_z^-))
\end{equation}
where the prefactor 24 is because there are 48 terms in the RHS of \eqref{eq:HEnv}.

Moreover, $[h_{e_z^\pm},\hat Q_v^{2n}]h_{e_z^\pm}^{-1}$ appearing in $\hat{H}_E^{(n)}(v;e_x^+,e_y^+,e_z^\pm)$ helps us to obtain the relation between the operators $\hat{H}_E^{(n)}(v;e_x^+,e_y^+,e_z^\pm)$. Indeed,
\begin{equation}\label{eq:ezpm}
\begin{aligned}
[h_{e^\pm_z},\hat Q_v^{2n}]h_{e^\pm_z}^{-1}=&\sum_{l=1}^{2n}\sum_{\mathcal P_l}\Big(\mp it\frac{(\beta a^2)^3}{8}\Big)^l\hat Q_v^{p_1}\epsilon_{\alpha_1\beta_1\gamma_1}\hat{X}^{\alpha_1}\hat{Y}^{\beta_1}\tau^{\gamma_1}\hat Q_v^{p_2}\epsilon_{\alpha_2\beta_2\gamma_2}\hat{X}^{\alpha_2}\hat{Y}^{\beta_2}\tau^{\gamma_2}
\cdots \hat Q_v^{p_l} \epsilon_{\alpha_l\beta_l\gamma_l}\hat{X}^{\alpha_l}\hat{Y}^{\beta_l}\tau^{\gamma_l}\hat Q_v^{p_{l+1}}
\end{aligned}
\end{equation}

where $e_z^\pm$ are oriented such that $\s(e_z^+)=v=\s(e_z^-)$, $\hat X^\alpha=\hat p_\s^\alpha(e')-\hat p_\t^\alpha(e')$ and $\hat Y^\alpha=\hat p_\s^\alpha(e'')-\hat p_\t^\alpha(e'')$, and $\mathcal P=\{p_1,p_2,\cdots,p_{l+1}\}$ with $p_i\in\mathbb Z$, $p_i\geq 0$ and $\sum_{i=1}^{l+1}p_i=2n-l$.\footnote{A general equation can be obtained analogously if the holonomy $h_{e_z^\pm}$ is replaced by a holonomy along other edges. In the following context, Eq. \eqref{eq:ezpm} will be usually referred as this general equation. }
Substituting the last equation, we have that
\begin{equation}\label{eq:HEezpezm}
\hat{H}_E^{(n)}(v;e_x^+,e_y^+,e_z^+)+\hat{H}_E^{(n)}(v;e_x^+,e_y^+,e_z^-)=2\hat{\tilde{H}}_E^{(n)}(v;e_x^+,e_y^+,e_z^+)
\end{equation}
where $\hat{\tilde{H}}_E^{(n)}(v;e_x^+,e_y^+,e_z^+)$ is the operator of $\hat{H}_E^{(n)}(v;e_x^+,e_y^+,e_z^+)$ with the replacement 
\begin{equation}\label{eq:ezpmreplacement}
\begin{aligned}
[h_{e_z^+},\hat Q_v^{2n}]h_{e_z^+}^{-1}\to& \sum_{l \text{ is odd} }\sum_{\mathcal P_l}\Big(- it\frac{(\beta a^2)^3}{8}\Big)^l\hat Q_v^{p_1}\epsilon_{\alpha_1\beta_1\gamma_1}\hat{X}^{\alpha_1}\hat{Y}^{\beta_1}\tau^{\gamma_1}\hat Q_v^{p_2}\epsilon_{\alpha_2\beta_2\gamma_2}\hat{X}^{\alpha_2}\hat{Y}^{\beta_2}\tau^{\gamma_2}
\cdots \hat Q_v^{p_l}\times\\ &\epsilon_{\alpha_l\beta_l\gamma_l}\hat{X}^{\alpha_l}\hat{Y}^{\beta_l}\tau^{\gamma_l}\hat Q_v^{p_{l+1}}.
\end{aligned}
\end{equation} 

According to Eq. \eqref{eq:HEnvp}, $\hat H_E^{(n)}(v)=48 \hat{\tilde{H}}_E^{(n)}(v;e_x^+,e_y^+,e_z^+)$. Thus
when calculating the expectation value of the Euclidean part, we only consider the operator $\hat{\tilde{H}}_E^{(n)}(v;e_x^+,e_y^+,e_z^+)$ instead of $\hat{H}_E^{(n)}(v;e_x^+,e_y^+,e_z^\pm)$. 

Moreover, Eq. \eqref{eq:ezpm} tells us that the Euclidean Hamiltonian takes the form $$\hat H_E^{(n)}(v;e_I,e_J,e_K)=\epsilon_{IJK}\tr(h_{\alpha_{IJ}}\tau^\alpha)\hat O_\alpha,$$
with $\hat O_\alpha$ a polynomial of fluxes. 
Because of $\tr(h\tau^\alpha)=-\tr(h^{-1}\tau^\alpha)$, we obtain that
\begin{equation}\label{eq:extrasymmtryofHE}
\hat H_E^{(n)}(v;e_I,e_J,e_K)=\hat H_E^{(n)}(v;e_J,e_I,e_K).
\end{equation}

In summary, originally there are 48 terms for each $\hat H_E^{(n)}(v)$ in Eq. \eqref{eq:HEnv}. Once we take into account the symmetries discussed in this section, we have
$\hat H_E^{(n)}(v)=48 \hat{\tilde{H}}_E^{(n)}(v;e_x^+,e_y^+,e_z^+)$. Thus we only need to compute the expectation value of the single term $\hat{\tilde{H}}_E^{(n)}(v;e_x^+,e_y^+,e_z^+)$.

\subsubsection{Symmetries of the Lorentzian part} 
Given a list of vertices and edges $(v;v_1,v_2,v_3,v_4;e_I,e_J,e_K)$ where $e_I,\ e_J$ and $e_K$ are outgoing from $v$, $(e_I, e_J, e_K)$ is either left-handed or right-handed. Thus
there exists a rotation $f$, leaving $v$ invariant, such that $(v;f(v_1),f(v_2),f(v_3),f(v_4);f(e_I),f(e_J),f(e_K))$ is either $(v;f(v_1),f(v_2),f(v_3),f(v_4);e_x^+,e_y^+,e_z^+)$ or $(v;f(v_1),f(v_2),f(v_3),f(v_4);e_x^+,e_y^+,e_z^-)$. Therefore, \eqref{eq:HLnv} can be simplified as
\begin{equation}\label{eq:HLnvp}
\hat H_L^{(\vec k)}(v)=24\sum_{v_1,v_2,v_3,v_4}\left(\hat{H}_L^{(\vec k)}(v;v_1,v_2,v_3,v_4;e_x^+,e_y^+,e_z^+)+\hat{H}_L^{(\vec k)}(v;v_1,v_2,v_3,v_4;e_x^+,e_y^+,e_z^-)\right)
 \end{equation}
Moreover, because the term $[h_{e_z^+},\hat Q_v^{2n}]h_{e_z^+}^{-1}$ also appears in $\hat H_L^{(\vec k)}$, Eq. \eqref{eq:ezpm} can be applied again which simplifies Eq. \eqref{eq:HLnvp} further to
\begin{equation}\label{eq:HLnvpp}
\hat H_L^{(\vec k)}(v)=48\sum_{v_1,v_2,v_3,v_4}\hat{\tilde H}_L^{(\vec k)}(v;v_1,v_2,v_3,v_4;e_x^+,e_y^+,e_z^+)
 \end{equation}
 where the $\hat{\tilde H}_L^{(\vec k)}$ operator is obtained from $\hat{H}_L^{(\vec k)}(v;v_1,v_2,v_3,v_4;e_x^+,e_y^+,e_z^+)$ by the replacement of Eq. \eqref{eq:ezpmreplacement}.  As a result, we only need to compute the expectation value of the operator $\hat{\tilde H}_L^{(\vec k)}(v;v_1,v_2,v_3,v_4;e_x^+,e_y^+,e_z^+)$ for various vertices $v_1,\, v_2,\, v_3$ and $v_4$.

The above discussion helps us to simplify our computation for the Lorentzian part. However, much more symmetries are needed so that the computation time is acceptable. 
For this purpose, let us firstly look at the term $[h_e[\hat Q_{v_1}^m,h_{e}^{-1}],\hat Q_{v_2}]$ which comes from the commutator between the volume operator and the Euclidean Hamiltonian. We obtain the following proposition which can be proven by using Eq. \eqref{eq:ezpm}. 

\begin{pro}\label{pro:simplifyQ}
Given an edge $e$ with the source $s(e)$ and the target $t(e)$, $
[h_e[\hat Q_{s(e)}^m,h_{e}^{-1}],\hat Q_{v}]= 0$ for all $v\neq s(e)$.
\end{pro}

With this proposition, let us consider the the commutator $ [\hat Q_{v_1}^{2k},\hat H_E^{(n)}(v_2;e_I,e_J,e_K)]$ which defines the operator $\hat K$. 
\begin{equation}\label{eq:quantumComHV}
\begin{aligned}
 &[\hat Q_{v_1}^{2k},\hat H_E^{(n)}(v_2;e_I,e_J,e_K)]\\
 =&\frac{1}{i\beta a^2 t}\epsilon_{IJK}\left(\tr([\hat Q_{v_1}^{2k},h_{\alpha_{IJ}}][h_{e_K},\hat Q_{v_2}^{2n}]h_{e_K}^{-1})+\tr(h_{\alpha_{IJ}}\big[\hat Q_{v_1}^{2k},[h_{e_K},\hat Q_{v_2}^{2n}]h_{e_K}^{-1}\big])\right)
\end{aligned}
\end{equation}
The commutators on the RHS of Eq. \eqref{eq:quantumComHV} can be expanded in $t$ with Eq. \eqref{eq:changeO} and simplified by the basic commutation relation \eqref{eq:commutators}, which can finally simplify $\hat H_L^{(\vec k)}(v)$ of Eq. \eqref{eq:HLnvpp} to be in terms of
\begin{equation}\label{eq:cruxofsimplification}
\frac{C}{t^2}\tr(h_{e_x^+}\, F_1 h_{e_x^+}^{-1}h_{e_y^+} F_2 h_{e_y^+}^{-1} G_1)
\end{equation}
with $C$ being some constant of order $t^0$ or higher,  $F_i$ being some monomials of holonomies and ($p_s^\alpha(e^+)\pm p_t^\alpha(e^-)$) and $G_1$ being a monomial of  ($p_s^\alpha(e^+)- p_t^\alpha(e^-)$). 

We apply the results in Sec. \ref{sec:leadingorder} to reduce the computational complexity. To use these results, we need to use the basic computation relations \eqref{eq:commutators} to simplify the Hamiltonian operator, so that the operators after the simplification are written in terms of  $C \hat P$ where $C$ is some constant of order $t^0$ or higher and $\hat P$ is some monomial of holonomies and fluxes.

In order to achieve so, we need to permute $h_{e_x^+}$ and $\hat F_1$, as well as $h_{e_y^+}$ and $\hat F_2$, in Eq. \eqref{eq:cruxofsimplification} by Eq. \eqref{eq:changeO}.
 Take permuting $h_{e_x^+}$ and $\hat F_1$ as an example:
Implementing the result of Eq. \eqref{eq:changeO}, we substitute $\hat O_1$ by $h_{e_x^+}$, and $\hat O_{i_k}$ by $\hat p_s^{\alpha}(e_x^+)$ and/or $\hat p_t^{\alpha}(e_x^+)$. One of many such substitutions inevitably generates some special terms whose commutators only contain $\hat p_s^{\alpha}(e_x^+)$. Computing these commutators with \eqref{eq:commutators} leads to results that are proportional to $h_{e_x^+}$. After substituting the permuted results into Eq. \eqref{eq:cruxofsimplification}, the $h_{e_x^+}$ eventually cancels with $h_{e_x^+}^{-1}$.
(similar to Eq. \eqref{eq:ezpm}). One can apply the same mechanism for permuting $h_{e_y^+}$ and $\hat F_2$.
 
 Let us collect these special terms from permuting $h_{e_x^+}$ and $\hat F_1$ and permuting $h_{e_y^+}$ and $\hat F_2$. We denote the partial sum of these special terms in $\hat H_L^{(k)}$ by ${}^{\rm alt}\hat{H}_L^{(\vec k)}$. Because of the cancellation between holonomies and their inverses discussed above, ${}^{\rm alt}\hat{H}_L^{(\vec k)}$ does not depend on $h_{e_x^+}$ and $h_{e_y^+}$. 
 
 It turns out that ${}^{\rm alt}\hat{H}_L^{(\vec k)}$ has more symmetries which are discussed shortly below. These special terms can be equivalently selected by considering only the non-commutativity between $h_{e_x^+}$ and $p_s^\alpha(e_x^+)$ while ignoring the non-commutativity between $h_{e_x^+}$ and $p_t^\alpha(e_x^+)$. Hence we have
 \begin{equation}\label{eq:halt0}
\text{the special terms of }h_{e_x^+}\hat F_1 h_{e_x^+}^{-1} =h_{s_x^+}\hat F_1 h_{s_x^+}^{-1}
 \end{equation}
 where $s_x^+$ is the segment within $e_x^+$ and does not contain the target $t(e_x^+)$. Note that the length of the segment does not cause any ambiguity because of the cancellation between the holonomies and their inverses discussed above. Indeed, Eq. \eqref{eq:halt0} leads to
\begin{equation}
\begin{aligned}
&{}^{\rm alt}\hat{H}_L^{(\vec k)}(v;v_1,v_2,v_3,v_4;e_I,e_J,e_K)\\
=&\frac{-2}{i\beta^7 a^{10} t^5}\epsilon^{IJK}\tr\left( [h_{s_I},[\hat Q_{v_1}^{2k_1},\hat{H}_E^{(k_2)}(v_2)]]h_{s_I}^{-1} [h_{s_J},[\hat Q_{v_3}^{2k_3},\hat{H}_E^{(k_4)}(v_4)]]h_{s_J}^{-1}[h_{s_K},\hat Q_v^{2k_5}]h_{s_K}^{-1}\right).
\end{aligned}
\end{equation}
Interestingly, the RHS may be understood as from an alternative definition of the Lorentzian part, 
\begin{equation}
\begin{aligned}
{}^{\rm alt}\widehat{H_L[N]}=\frac{-2}{i\beta^7 a^{10} t^5}\sum_v N(v)\sum_{s_I,s_J,s_K \text{ at } v}\varepsilon^{IJK}\tr\left( [h_{s_I},[\hat V,\hat H_E]]h_{s_I}^{-1} [h_{s_J},[\hat V,\hat H_E]]h_{s_J}^{-1}[h_{s_K},\hat V_v]h_{s_K}^{-1}\right).\label{Halt}
\end{aligned}
\end{equation}
where all edges $e_I,e_J,e_K$ are replaced by the segments $s_I,s_J,s_K$ with $s_I\subset e_I$. ${}^{\rm alt}\widehat{H_L[N]}$ may be obtained by a different regularization/quantization, 
$$\{K,\dot{e}^aA_a(x)\}\to \frac{-1}{2\kappa(i\hbar\beta)^2}[h_{s_e},[\hat{V},\hat{H}_E]] h_{s_e}^{-1},$$
with $\dot e^a$ the vector tangent to the edge $e$ and $s_e\subset e$ the segment. More details on this aspect is discussed in \cite{toappear}.

We collect terms in $\hat H_L^{(k)}$ that are other than the special terms discussed above, and denote their sum by ${}^{\rm extra}\hat{H}_L^{(\vec k)}$, namely
\begin{eqnarray}
&&{}^{\rm extr}\hat{H}_L^{(\vec k)}(v;v_1,v_2,v_3,v_4;e_I,e_J,e_K)\nonumber\\
&=&\hat{H}_L^{(\vec k)}(v;v_1,v_2,v_3,v_4;e_I,e_J,e_K)-{}^{\rm alt}\hat{H}_L^{(\vec k)}(v;v_1,v_2,v_3,v_4;e_I,e_J,e_K).\label{altextra}
\end{eqnarray}
The operators ${}^{\rm alt}\hat{H}_L^{(\vec k)}$ and ${}^{\rm extr}\hat{H}_L^{(\vec k)}$ are treated separately in our algorithm.   

For ${}^{\rm alt}\hat{H}_L^{(\vec k)}$, the above simplification procedures lead to 
\begin{equation}\label{eq:cruxsimplificationsolved}
\frac{C}{t^2}h_{s_x^+}\, F_1 h_{s_x^+}^{-1}h_{s_y^+} F_2 h_{s_y^+}^{-1} G_1,
\end{equation}
instead of Eq. \eqref{eq:cruxofsimplification}. 
Since $[h_{s_I},\hat p_\t^\alpha(e_I)]=0$, these terms can be simplified with
\begin{equation}\label{eq:hphm1}
\begin{aligned}
h_{\s^\pm}\prod_{i=1}^m (\sigma^+_ip^{\alpha_i}_\s(e^+)+\sigma^-_i p^{\alpha_i}_\t(e^-))h_{s^\pm}^{-1}=\sum_{\mathcal I}(i t)^{|\mathcal I|}\prod_{i\notin \mathcal I}(\sigma_i^+p^{\alpha_i}_\s(e^+)+\sigma_i^- p^{\alpha_i}_\t(e^-))\prod_{j\in\mathcal I}\sigma_j^\pm \tau^{\alpha_j}
\end{aligned}
\end{equation}
where $\mathcal I$ is a subsets of $\{1,2,\cdots,m\}$ with length $|\mathcal I|$, $s^+$ and $s^-$ are segments of $e^+$ and $(e^-)^{-1}$ respectively, $\sigma_i^+=1$ and  $\sigma_i^-=\pm1$. Because of the summation over $e_I$, $e_J$ and $e_K$ in Eq. \eqref{eq:HLnv} which lets us to compute 
\begin{equation}
h_{s^+}\prod_{i=1}^m (\sigma^+_ip^{\alpha_i}_\s(e^+)+\sigma^-_i p^{\alpha_i}_\t(e^-))h_{s^+}^{-1}-h_{s^-}\prod_{i=1}^m (\sigma^+_ip^{\alpha_i}_\s(e^+)+\sigma^-_i p^{\alpha_i}_\t(e^-))h_{s^-}^{-1},
\end{equation} 
we can use the replacement
\begin{equation}\label{eq:hphm1replacement}
h_{s^+}\prod_{i=1}^m (\sigma^+_ip^{\alpha_i}_\s(e^+)+\sigma^-_i p^{\alpha_i}_\t(e^-))h_{s^+}^{-1}\to \sum_{\mathcal J}(i t)^{|\mathcal J|}\prod_{i\notin \mathcal J}(\sigma_i^+p^{\alpha_i}_\s(e^+)+\sigma_i^- p^{\alpha_i}_\t(e^-))\prod_{j\in\mathcal J} \tau^{\alpha_j}
\end{equation}
with $\mathcal J\subset\{1,2,\cdots,m\}$ such that 
\begin{equation}
\prod_{j\in \mathcal J}\sigma_j^-=-1.
\end{equation}

Substituting Eq. \eqref{eq:hphm1replacement} into \eqref{eq:cruxsimplificationsolved}, we cancel the prefactor $1/t^2$ to simplify  \eqref{eq:cruxsimplificationsolved} to be in terms of $C\hat P$ with some constant $C$ of order $t^0$ or higher, so that the results in Sec. \ref{sec:leadingorder} can be applied. 

Moreover, ${}^{\rm alt}\hat{H}_L^{(\vec k)}$ brings more symmetries, which is described as below: Consider a $\pi$-rotation which transforms either $e_x^+$ to $e_x^-$ or $e_y^+$ to $e_y^-$. Denote this rotation by $\mathfrak s$. Let us rewrite $F_i$ in Eq. \eqref{eq:cruxsimplificationsolved} as $F_i(v,e)$ to indicate the dependence of $F_i$ on vertices and edges. Then Eq. \eqref{eq:hphm1} tells us 
\begin{equation}
\begin{aligned}
&h_{\mathfrak{s}(s_x^+)}F_1(\mathfrak s(v),\mathfrak s(e))h_{\mathfrak{s}(s_x^+)}^{-1}h_{\mathfrak{s}(s_y^+)}F_2(\mathfrak s(v),\mathfrak s(e))h_{\mathfrak{s}(s_y^+)}^{-1}G_1(\mathfrak s(v),\mathfrak s(e))\\
=&h_{s_x^+}F_1(\mathfrak s(v),\mathfrak s(e))h_{s_x^+}^{-1}h_{s_y^+}F_2(\mathfrak s(v),\mathfrak s(e))h_{s_y^+}^{-1}G_1(\mathfrak s(v),\mathfrak s(e)).
\end{aligned}
\end{equation}
Furthermore, recalling Eq. \eqref{eq:DifftoOp}, we obtain that
\begin{equation}\label{eq:symmetryHalt}
\begin{aligned}
&\langle h_{s_x^+}F_1(v,e)h_{s_x^+}^{-1}h_{s_y^+}F_2(v,e)h_{s_y^+}^{-1}G_1(v,e)\rangle\\
=&\langle h_{s_x^+}F_1(\mathfrak s(v),\mathfrak s(e))h_{s_x^+}^{-1}h_{s_y^+}F_2(\mathfrak s(v),\mathfrak s(e))h_{s_y^+}^{-1}G_2(\mathfrak s(v),\mathfrak s(e))\rangle
\end{aligned}
\end{equation}
which reduces the number of contributing vertices and edges in the computation of ${}^{\rm alt}\hat{H}_L^{(\vec k)}$.

For the operator ${}^{\rm extr}\hat{H}_L^{(\vec k)}$, Eq. \eqref{eq:hphm1} can no longer be used and, thus, there does not exist the symmetry implied by \eqref{eq:symmetryHalt} yet. We propose the following strategy to reduce the complexity of the computation. Let $\mathfrak r$ be the rotation about the axis $(1/\sqrt{2},1/\sqrt{2},0)$ for $\pi$ radians, which exchanges the $x$- and $y$-axes, and flips the $z$-axis. We have
\begin{equation}\label{eq:trotatedHL}
\begin{aligned}
\hat{\tilde H}_L^{(\vec k)}(v;v_1,v_2,v_3,v_4;e_x^+,e_y^+,e_z^+)=\hat{\tilde H}_L^{(\vec k)}(v;\mathfrak r(v_1),\mathfrak r(v_2),\mathfrak r(v_3),\mathfrak r(v_4);e_y^+,e_x^+,e_z^+)
\end{aligned}
\end{equation}
where $\hat{\tilde H}_L^{(\vec k)}$ is defined in  \eqref{eq:HLnvpp} and $\mathfrak r(v)=v$ is assumed without loss of generality. Let us consider the operator 
\begin{equation}
\begin{aligned}
&\hat F(v;v_1,v_2,v_3,v_4;e_x^+,e_y^+,e_z^+)\\
:=&\hat{H}_L^{(\vec k)}(v;v_1,v_2,v_3,v_4;e_x^+,e_y^+,e_z^+)+\hat{H}_L^{(\vec k)}(v;\mathfrak r(v_3),\mathfrak r(v_4),\mathfrak r(v_1),\mathfrak r(v_2);e_x^+,e_y^+,e_z^+)
\end{aligned}
\end{equation}
 By \eqref{eq:trotatedHL}, $\hat F$, up to an overall factor, is 
 \begin{equation}
\begin{aligned}
&\hat F(v;v_1,v_2,v_3,v_4;e_x^+,e_y^+,e_z^+)\\
=&-\tr( [h_{e_y^+},[\hat Q_{\mathfrak r(v_1)}^{2k_1},\hat{H}_E^{(k_2)}(\mathfrak r(v_2))]]h_{e_y^+}^{-1} [h_{e_x^+},[\hat Q_{v_3}^{2k_3},\hat{H}_E^{(k_4)}(\mathfrak r(v_4))]]h_{e_x^+}^{-1}[h_{e_z^+},\hat Q_v^{2k_5}]h_{e_z^+}^{-1})\\
&+\tr( [h_{e_x^+},[\hat Q_{\mathfrak r(v_3)}^{2k_1},\hat{H}_E^{(k_2)}(\mathfrak r(v_4))]]h_{e_x^+}^{-1} [h_{e_y^+},[\hat Q_{\mathfrak r(v_1)}^{2k_3},\hat{H}_E^{(k_4)}(\mathfrak r(v_2))]]h_{e_y^+}^{-1}[h_{e_z^+},\hat Q_v^{2k_5}]h_{e_z^+}^{-1})
\end{aligned}
\end{equation}
For clarity, let us denote 
\begin{equation}
\begin{aligned}
\hat X\equiv &[h_{e_x^+},[\hat Q_{v_3}^{2k_3},\hat{H}_E^{(k_4)}(\mathfrak r(v_4))]]h_{e_x^+}^{-1}\\
\hat Y\equiv  &[h_{e_y^+},[\hat Q_{\mathfrak r(v_1)}^{2k_1},\hat{H}_E^{(k_2)}(\mathfrak r(v_2))]]h_{e_y^+}^{-1}\\
\hat Z\equiv &[h_{e_z^+},\hat Q_v^{2k_5}]h_{e_z^+}^{-1}.
\end{aligned}
\end{equation}
Because of the holonomies in them, they are all operator-valued matrices whose entries are denoted as $\hat X_{ab}$, $\hat Y_{ab}$ and $\hat Z_{ab}$ respectively. With this notion, we have
\begin{equation}
\hat F=\hat X_{ab}\hat Y_{bc}\hat Z_{ca}-\hat Y_{ab}\hat X_{bc}\hat Z_{ca}=\hat Y_{bc}\hat X_{ab}\hat Z_{ca}-\hat Y_{ab}\hat X_{bc}\hat Z_{ca}+[\hat X_{ab},\hat Y_{bc}]\hat Z_{ca}.\label{FXYZ}
\end{equation}
For the first two terms, because of Eq. \eqref{eq:basicformula}, the expectation value of their subtraction can be simplified as
\begin{equation}
\begin{aligned}
\langle \hat Y_{bc}\hat X_{ab}\hat Z_{ca}-\hat Y_{ab}\hat X_{bc}\hat Z_{ca}\rangle_{\{g_e\}}=(\mathcal C_{\tilde a\tilde b\tilde c\tilde d\tilde e\tilde f}-\mathcal C'_{\tilde a\tilde b\tilde c\tilde d\tilde e\tilde f})\langle \hat{Y'}_{\tilde a\tilde b}\hat{X'}_{\tilde c \tilde d}\hat{Z'}_{\tilde e \tilde f}\rangle_{\{z_e\}}
\end{aligned}
\end{equation}
where $\mathcal C_{\tilde a\tilde b\tilde c\tilde d\tilde e\tilde f}$ and $\mathcal C'_{\tilde a\tilde b\tilde c\tilde d\tilde e\tilde f}$ are two sets of constant coefficients. $(\mathcal C_{\tilde a\tilde b\tilde c\tilde d\tilde e\tilde f}-\mathcal C'_{\tilde a\tilde b\tilde c\tilde d\tilde e\tilde f})$ can be computed explicitly. $\hat X'$, $\hat Y'$ and $\hat Z'$ are the operators gauge transformed from $X,Y,Z$ according to Eq. \eqref{eq:basicformula}. For the last term in \eqref{FXYZ}, since $[\hat X_{ab},\hat Y_{bc}]$ itself is $O(t)$, we only need to compute the leading order expectation value of this term. Thus  the commutator $[\hat X_{ab},\hat Y_{bc}]$ can be easily computed with the results in Sec. \ref{sec:leadingorder}. 

Let us summarize the effects of symmetries. Let $N_1$ be the original number of terms in ${}^{\rm alt}\hat H_L^{(k)}$. We firstly use the symmetries implied by Eqs. \eqref{eq:HEezpezm} and \eqref{eq:extrasymmtryofHE} to reduce the number to $N_1/4^2$, where the power 2 is because there are two $\hat H_E$ in ${}^{\rm alt}\hat H_L^{(k)}$. Eq. \eqref{eq:HLnvpp} further reduces this number to $N_1/(4^2\times 48)$. Finally the symmetry \eqref{eq:symmetryHalt} reduces the number to $$\frac{N_1}{4^2\times 48\times 4}= \frac{N_1}{3072}.$$
Let $N_2$ be the original number of terms in ${}^{\rm extr}\hat H_L^{(k)}$. We firstly use the symmetries implied by Eqs. \eqref{eq:HEezpezm} and \eqref{eq:extrasymmtryofHE} to reduce the number to $N_2/4^2$. Then Eq. \eqref{eq:HLnvpp} reduces the number to $N_2/(4^2\times 48)$. Eq. \eqref{FXYZ} further reduce this number to about \footnote{The word ``about" is because there exist the cases with $\hat X=\hat Y$ in Eq. \eqref{FXYZ}.}
$$\frac{N_2}{4^2\times 48\times 2}= \frac{N_2}{1536}.$$

\subsection{Introducing the algorithm: A case study} \label{Introducing the algorithm: A case study}

The procedure to compute the expectation value of the Hamiltonian operator can be illustrated by the following simple example. The complete implementation is presented in \cite{toappear}. Let $\hat O$ be the operator 
\begin{equation}
\begin{aligned}
\hat O=\frac{1}{t}[D^\iota_{ab}(h_{e_z^+}),\hat Q_v^{m}]D^\iota_{b c}(h_{e_z^+}^{-1})
\end{aligned}
\end{equation} 
where $v=s(e_z^+)$. Then, in the case when we are only interested in  $O(t)$ expansion,  the expectation value of $\hat O$ can be simplified by Eq. \eqref{eq:ezpm},
\begin{equation}
\begin{aligned}
\langle \hat O\rangle_{g}=&\Big( i\frac{(\beta a^2)^3}{8}\Big)\sum_{k}\left(\prod_{i=1}^m\mathcal T_{\alpha_i\beta_i\gamma_i}\right)D^\iota_{ac}(\tau^{\gamma_k})\left\langle\prod_{i=1}^m\hat X^{\alpha_i}\right\rangle_{z}\left\langle\prod_{i=1}^m\hat Y^{\beta_i}\right\rangle_{z}\left\langle\prod_{i\neq k}\hat Z^{\gamma_i}\right\rangle_z\\
&+\Big( i\frac{(\beta a^2)^3}{8}\Big)^2 t\sum_{k<l}\left(\prod_{i=1}^m\mathcal T_{\alpha_i\beta_i\gamma_i}\right)D^\iota_{ac}(\tau^{\gamma_k}\tau^{\gamma_l})\left\langle\prod_{i=1}^m\hat X^{\alpha_i}\right\rangle_{z}\left\langle\prod_{i=1}^m\hat Y^{\beta_i}\right\rangle_{z}\left\langle\prod_{i\neq k,l}\hat Z^{\gamma_i}\right\rangle_z\\
&=:O_0+t O_1
\end{aligned}
\end{equation}
where $\mathcal T_{\alpha\beta\gamma}$ is the contraction of $\epsilon_{\alpha'\beta'\gamma'}$ with the gauge transformation matrices according to Eq. \eqref{eq:basicformula}. 

Now let us show how to apply the results in Sec. \ref{sec:leadingorder} to reduce the computation complexity.  Take $O_0$ as an example. 
The result of $O_0$ can be divided into two parts: the expectation-value parts 
$$\left\langle\prod_{i=1}^m\hat X^{\alpha_i}\right\rangle_{z}\left\langle\prod_{i=1}^m\hat Y^{\beta_i}\right\rangle_{z}\left\langle\prod_{i\neq k}\hat Z^{\gamma_i}\right\rangle_z$$
and the factor parts
$$\left(\prod_{i=1}^m\mathcal T_{\alpha_i\beta_i\gamma_i}\right)D^\iota_{ac}(\tau^{\gamma_k}).$$
For the expectation-value parts, without the results given in Sec. \ref{sec:leadingorder}, we would have to consider $3^{3m-1}$ cases for various values of the indices $\alpha_i$, $\beta_i$ and $\gamma_i$. However, once the results in Sec. \ref{sec:leadingorder} are applied, since we only expand the expectation value to the next-to-leading order, we only have the following options
\begin{itemize}
\item[(1)] all indices are $0$;
\item[(2)] these indices contain a single $+1$  or a single $-1$; 
\item[(3)] these indices contain a pair of $(+1,-1)$  or a pair $(-1,+1)$. 
\end{itemize} 
Moreover, the results in Sec. \ref{sec:leadingorder} also tell us that for the last two options, positions of the $\pm 1$ among these indices do not influence the expectation value, as far as the leading and next-to-leading orders are concerned. Therefore, we finally only need to compute $1+2+2=5$ cases. Comparing with the original number of cases $3^{3m-1}$, we conclude that the computational complexity is reduced exponentially by the algorithm. 

For the factor part, we only need to combine $\gamma_k$ of $D^\iota_{ac}(\tau^{\gamma_k})$ with the indices in the expectation-value part. Then we consider all possibilities of permutations and all values of $\gamma_k$ so that all possible values of the factor part can be computed. The result of $O_0$ can be obtained by summing over all possibilities.

Finally, let us complete this section with discussing the values of $n$ and $\vec k$ in $\hat H_E^{(n)}$ and $\hat H_L^{(\vec k)}$. When we replace $\hat{V}_v$ in $\widehat{H_E}$ by $\hat{V}_{G T}^{(v)}$, a term $\left(\hat Q^2/\langle\hat Q\rangle^2-1\right)^k$ in $\hat{V}_{G T}^{(v)}$ has the following contribution to $\widehat{H_E}$
\begin{equation}\label{eq:operatorqmqk}
\begin{aligned}
&\frac{1}{t}\tr(h_{\alpha_{IJ}}h_{s_K}[\left(\hat Q^2/\langle\hat Q\rangle^2-1\right)^k,h_{s_K}^{-1}])\\
=&\sum_{m=0}^{k-1}\tr(h_{\alpha_{IJ}}h_{s_K}\left(\hat Q^2/\langle\hat Q\rangle^2-1\right)^m\frac{[\hat Q^2/\langle\hat Q\rangle^2,h_{s_K}^{-1}]}{t}\left(\hat Q^2/\langle\hat Q\rangle^2-1\right)^{k-1-m})
\end{aligned}
\end{equation}
where the unimportant overall coefficient is neglected. Because $\hat Q^2/\langle\hat Q\rangle^2-1=\left(\hat Q/\langle\hat Q\rangle+1\right)\left(\hat Q/\langle\hat Q\rangle-1\right)$ and the expectation value of $ \hat Q/\langle\hat Q\rangle-1$ is order $O(t)$, Eq. \eqref{thm:leadingordergeneral} tells us that the leading order of the operator in Eq. \eqref{eq:operatorqmqk} is of $O(t^{\floor{k/2}})$. Thus as far as expanding the expectation value to $O(t)$ is considered, it is sufficient to choose $k\leq 3$ which means that the value of $n$ in $\hat H_E^{(n)}$ is smaller than or equal to $3$. A very similar discussion for $\hat H_L^{(\vec k)}$ gives us that $\vec k$ satisfies
\begin{equation}
\frac{|k_1+k_2-3|+(k_1+k_2-3)}{2}+\frac{|k_3+k_4-3|+(k_3+k_4-3)}{2}+k_5\leq 3.
\end{equation}

\section{Quantum correction in the expectation value} \label{sec:results}

We summarize the resulting expectation value of the Hamiltonian with unit lapse $\widehat{H[1]}=\widehat{H_{E}}+\left(1+\beta^{2}\right) \widehat{H_{L}}$ at coherent states with cosmological data ($\eta<0$ in our convention)
\begin{eqnarray}
\langle \widehat{H_E}\rangle&=&6 a \sqrt{-\beta  \eta } \sin ^2(\xi )-\frac{3}{4} a t \sqrt{-\frac{\beta }{\eta ^3}} \sin ^2\left(\frac{\xi }{2}\right) \cos\left(\frac{\xi }{2}\right) \Bigg\{\cos\left(\frac{\xi }{2}\right) \Big[8 \eta ^2+8 \eta  (4 \cosh (\eta )-3) \text{csch}(\eta )-9\Big]\nonumber\\
&&-\,12 i \eta  \sin \left(\frac{\xi }{2}\right)\Bigg\}+O(t^2),\label{euclidhamres}\\
\langle \widehat{H_L}\rangle&=&-\frac{6 a \sqrt{-\beta  \eta } \sin ^2(\xi ) \cos ^2(\xi )}{\beta ^2}-\frac{3 a t}{262144 (-\beta  \eta )^{3/2}} \Bigg\{2 \left(3-220 \eta ^2\right) \cos (6 \xi )\nonumber\\
&&+\,4 i \eta  (4838 \sin (\xi )-6284 \sin (2 \xi )+4685 \sin (3 \xi )-5222 \sin (4 \xi )-105 \sin (5 \xi ))\nonumber\\
&& +\,2 (-3611+8 \eta  (492 \eta +11 i)) \cos (\xi )-2 (-789+4 \eta  (305 \eta +18 i)) \cos (2 \xi )\nonumber\\
&& +\,(4413-4 \eta  (928 \eta +49 i)) \cos (3 \xi )+8 (-1978+\eta  (4192 \eta -7 i)) \cos (4 \xi )\nonumber\\
&& +\,(-7+4 (-272 \eta +5 i) \eta ) \cos (5 \xi )-4 \eta  \coth (\eta ) \Big[536 \cos (\xi )+1731 \cos (2 \xi )\nonumber\\
&&+\,1524 \cos (3 \xi )-40548 \cos (4 \xi )+116 \cos (5 \xi )+117 \cos (6 \xi )+37292\Big]\nonumber\\
&&+\,8 \eta  \text{csch}(\eta ) \Big[130 \cos (\xi )+918 \cos (2 \xi )+801 \cos (3 \xi )-18618 \cos (4 \xi )\nonumber\\
&&+\,125 \cos (5 \xi )+58 \cos (6 \xi )+16362\Big]+8 (1436+\eta  (-4056 \eta +25 i))\Bigg\}+O(t^2). 
\end{eqnarray}
By \eqref{altextra}, $\langle \widehat{H_L}\rangle=\langle{}^{\rm extr}\widehat{H_L}\rangle+\langle{}^{\rm alt}\widehat{H_L}\rangle$ where $\langle{}^{\rm alt}\widehat{H_L}\rangle$ is the expectation value of \eqref{Halt}, 
\begin{equation}
\begin{aligned}
\langle{}^{\rm alt}\widehat{H_L}\rangle=&-\frac{3 a \sqrt{-\beta  \eta } \sin ^2(2 \xi )}{8 \beta ^2}-\frac{3 a t}{32768 (-\beta  \eta )^{3/2}}\Bigg\{4 \left(104 \eta ^2-79\right) \cos (\xi )+\left(68-160 \eta ^2\right) \cos (2 \xi )\\
&+4 \left(55-104 \eta ^2\right) \cos (3 \xi )+\left(944 \eta ^2-501\right) \cos (4 \xi )-848 \eta ^2\\
&-2 \eta  \Big[\coth (\eta ) (-(262 \cos (\xi )-6 (46 \cos (2 \xi )+65 \cos (3 \xi )-380 \cos (4 \xi )+366)))\\
&-\text{csch}(\eta ) (-292 \cos (\xi )+268 \cos (2 \xi )+420 \cos (3 \xi )-2035 \cos (4 \xi )+1799)\\
&-64 i (6 \sin (\xi )-10 \sin (2 \xi )+6 \sin (3 \xi )-7 \sin (4 \xi ))\Big]+241\Bigg\}+O(t^2),
\end{aligned}
\end{equation}
\begin{equation}
\begin{aligned}
\langle{}^{\rm extr}\widehat{H_L}\rangle=&-\frac{9 a \sqrt{-\beta  \eta } \sin ^2(2 \xi )}{8 \beta ^2}+\frac{3 a t }{262144 (-\beta  \eta) ^{3/2}} \Bigg\{-2 \left(580 \eta ^2+72 i \eta -517\right) \cos (2 \xi )\\
&+2 \left(3-220 \eta ^2\right) \cos (6 \xi )+4 i \eta  \Big[3302 \sin (\xi )-3724 \sin (2 \xi )+3149 \sin (3 \xi )\\
&-35 (98 \sin (4 \xi )+3 \sin (5 \xi ))\Big]+2 (-2347+8 \eta  (284 \eta +11 i)) \cos (\xi )\\
&+(2653-4 \eta  (96 \eta +49 i)) \cos (3 \xi )+56 (-211+\eta  (464 \eta -i)) \cos (4 \xi )\\
&+(-7+4 (-272 \eta +5 i) \eta ) \cos (5 \xi )-4 \eta  \coth (\eta ) \Big[1584 \cos (\xi )+627 \cos (2 \xi )\\
&-36 \cos (3 \xi )-31428 \cos (4 \xi )+116 \cos (5 \xi )+117 \cos (6 \xi )+28508\Big]\\
&+8 \eta  \text{csch}(\eta ) \Big[714 \cos (\xi )+382 \cos (2 \xi )-39 \cos (3 \xi )\\
&-14548 \cos (4 \xi )+125 \cos (5 \xi )+58 \cos (6 \xi )+12764\Big]+8 \eta  (-3208 \eta +25 i)+9560\Bigg\}+O(t^2).
\end{aligned}
\end{equation}

 {Note that the above results are written based on the graph with only a single vertex. Moreover, the results are complex because the operators $\widehat{H_E}$ and $\widehat{H_L}$ are not symmetry operators. Usually, we define the symmetrized operator as $\frac12(\widehat{H_E}+\widehat{H_E}^\dagger)$ and $\frac12 (\widehat{H_L}+\widehat{H_L}^\dagger)$. Then the expectation values 
$\langle \frac12(\widehat{H_E}+\widehat{H_E}^\dagger)\rangle$, and $\langle\frac12 (\widehat{H_L}+\widehat{H_L}^\dagger)\rangle$ are just the real parts of $\langle\widehat{H_E}\rangle$ and $\langle\widehat{H_L}\rangle$ respectively. Furthermore, }
when we express $\langle \widehat{H[1]}\rangle=H_0+t H_1+O(t^2)$, we notice that the $O(t)$ contribution $H_1$ contains $\eta$ in the denominator, thus is divergent if $\eta\to0$. This feature descends from the fact that $\hat{V}_{GT}^{(v)}$ is divergent if $\langle \hat{Q}_v\rangle\to0$ because $\langle \hat{Q}_v\rangle\sim |\eta|^{3}$. Thus our expansion of $\langle \widehat{H[1]}\rangle$ becomes invalid if $\eta$ is too small. More precisely, when we express $\langle \widehat{H[1]}\rangle=\sqrt{|\eta|}\left[\mathfrak{f}_0+({t}/{\eta^2})\, \mathfrak{f}_1+O(t^2)\right]$, we have that $\mathfrak{f}_0$ is independent of $\eta$, and $\mathfrak{f}_1$ is regular at $\eta\to0$, so we conclude that our expansion is valid when $ \eta^{2}\gg t $. Our expansion is valid for large $|\eta|$, because $\mathfrak{f}_1$ is regular at $|\eta|\to\infty$.

 {Even though there have been literature (e.g. \cite{Dapor:2017rwv,Dapor:2017gdk,Liegener:2019jhj}) considering the same issue and telling the leading order of the expectation value, our work gives the $O(\hbar)$ correction of the both Euclidean part and Lorentzian part. It should be noted that the  $O(t)$ correction presented in the work is just the correction to the Hamiltonian operator. To get the total quantum correction of the whole mode, i.e. the total quantum correction of the effective action, one also needs to consider the 1-loop correction as in the usual quantum field theory. 
}

 {
 Finally, we would like to compare our results with the results in loop quantum cosmology. In loop quantum cosmology, there are two types of dynamics, i.e., the $\mu_0$-scheme dynamics and the $\bar\mu$-scheme dynamics  \cite{ashtekar2006quantumnatureI,ashtekar2006quantumnature}. Since the $\mu_0$-scheme dynamics possesses some physically undesirable features, saying, that the bounce could occur in some classical regime. People prefer the $\bar\mu$-scheme dynamics solving those problems in the $\mu_0$-scheme model. However, according to our results, our work can only recover the $\mu_0$ scheme of loop quantum cosmology, and  further adds some incomplete $O(t)$ corrections for cosmology. It is still open whether one can recover the $\bar\mu$-scheme dynamics in the current framework or not once all quantum corrections are considered. Beyond the current framework, another attempt to recover the $\bar\mu$-scheme cosmology is to considering the graph-changing Hamiltonian operator. For instance, in \cite{han2021loop}, by considering a model where the lattice changes in time, one gets results qualitatively similar to the $\bar\mu$-scheme model.
 }

\section{Conclusion and outlook}\label{sec:conclusion}

In this paper, we develop an algorithm to overcome the complexity in the LQG Hamiltonian operator $\widehat{H[N]}$, and we compute explicitly the $O(\hbar)$ quantum correction in the expectation value $\langle\widehat{H[N]}\rangle$ at the coherent state peaked at the homogeneous and isotropic data of cosmology. With our algorithm and results, there are several perspectives which should be addressed in the future:

An important future task is to complete the computation of the quantum effective action $\Gamma$ mentioned in Section \ref{sec:intro}. After the completion of the present work, the only missing ingredient in the $O(\hbar)$ terms in $\Gamma$ is the ``1-loop determinant'' $\det(\mathfrak{H})$. Therefore a research to be carried out immediately is to compute $\det(\mathfrak{H})$ at the homogeneous and isotropic background. Once we obtain the complete $O(\hbar)$ contribution to $\Gamma$, the variation of $\Gamma$ should give the quantum corrected effective equations which demonstrate the quantum correction to cosmology implied by LQG.    

The next step of generalizing our computation is to study the expectation values $\langle\widehat{H[N]}\rangle$ at coherent states peaked at cosmological perturbations. The semiclassical limit of the expectation value and the cosmological perturbation theory from the path integral \eqref{Agg0} are studied in \cite{Han:2020iwk}. Thus it is interesting to compute the $O(\ell_{\rm p})$ correction to the cosmological perturbation theory.

The computation of the quantum correction in $\langle\widehat{H[N]}\rangle$ should be extended to couplings of standard matter fields. The contributions from matter fields to $\widehat{H[N]}$ have been studied in the literature \cite{Thiemann:1997rt,Sahlmann:2002qj}. The matter parts in $\widehat{H[N]}$ is much simpler than the Lorentzian part in $\widehat{H[N]}$, so computing their expectation values should not be hard. Studying the matter contributions and their quantum corrections is a project currently undergoing \cite{inpreparation}.

\section*{Acknowledgements}

M.H. acknowledges Andrea Dapor, Klaus Liegener, and Hongguang Liu for various discussions motivating this work. M.H. receives support from the National Science Foundation through grant PHY-1912278. C. Z. acknowledges the support by the Polish Narodowe Centrum Nauki, Grant No. 2018/30/Q/ST2/00811

\bibliographystyle{jhep}

\bibliography{reference}

\end{document}